\documentclass[12pt]{article}

\usepackage{tikz}
\usepackage{epsfig,latexsym,amsfonts,amsmath,amsthm,amssymb,amsbsy,multirow,slashed,wasysym,textcomp,subfigure,wrapfig,datetime,comment,mathtools,cancel,cite,mathrsfs}
\usepackage[hidelinks]{hyperref}

\def\ee{\end{equation}}
\def\be{\begin{equation}}
\def\bea{\begin{eqnarray}}
\def\eea{\end{eqnarray}}
\newcommand{\beq}{\begin{eqnarray}}
\newcommand{\eqq}{\end{eqnarray}}
 \newcommand{\badat}{\begin{alignedat}}
 \newcommand{\eadat}{\end{alignedat}}

\newcommand{\eal}[1]{\be \begin{aligned} #1 \end{aligned}\end{equation}} 
\newcommand{\eqn}[1]{\be #1 \end{equation}} 
\newcommand{\eqa}[1]{\bea  #1\end{eqnarray}}

\long\def\new#1\endnew{{\bf #1}}		
\long\def\del#1\enddel{}

\def\del{\partial}

\usepackage{color}
\usepackage{wrapfig}
\usepackage{MnSymbol}

\definecolor{orange}{rgb}{1,0.5,0}

\newcommand{\pink}[1]{\textcolor{\pink}{#1}}

\definecolor{dblue}{rgb}{0.2,0.50,0.80}

\def\et2{ET$^2$}
\def\m2{M$^2$}

\def\bz{{\bar z}}

\numberwithin{equation}{section} 

\begin{document}
\begin{titlepage}
  \thispagestyle{empty}
  \begin{flushright}
    \end{flushright}
  \bigskip
  \begin{center}
	 \vskip2cm
  \baselineskip=13pt {\LARGE Soft Algebra for ${\cal N}=4$ SYM  
}

	 \vskip2cm
   \centerline{Luis F. Alday$^a$ and Andrew Strominger$^{b,c}$}
    \vskip.8cm
 \noindent{\em $^a$University of Oxford, 
  $^b$OpenAI,
$^c$Harvard University}

\bigskip
  \vskip1cm
  \end{center}

  \begin{abstract}
 Scattering amplitudes of $n$ particles in nonabelian gauge theories admit factorizations of the general form $\mathcal{A}_n \;=\; \mathcal{A}^{\rm soft}_n \times \mathcal{A}^{\rm hard}_n$, where $\mathcal{A}^{\rm soft}_n$ is IR divergent, while $\mathcal{A}^{\rm hard}_n$ is IR finite {and encodes the higher loop corrections to scattering}. We specify a particular all-orders definition of this factorization for planar ${\cal N}=4$ super Yang-Mills (SYM) and argue that the resulting $\mathcal{A}_n^{\rm hard}$ obeys an uncorrected tree-level soft theorem. Moreover it furnishes a representation of the undeformed tree-level $\cal S$-algebra generated by a tower of soft gluons. The results follow from several commonly invoked assumptions for ${\cal N}=4$ SYM, including  BDS one-loop exponentiation of the splitting function and amplitude/Wilson-loop duality.    \end{abstract}
%

%
\end{titlepage}
\tableofcontents
\section{Introduction}

Classical nonabelian gauge theory and gravity are both governed in (asymptotically) flat space by infinite-dimensional soft symmetry algebras, known as the $S$-algebra and ${\cal L}w_{1+\infty}$-algebra, respectively, which act primarily on the $\cal S$-matrix. These can be understood variously as Ward identities of soft theorems, asymptotic symmetries, chiral algebras of D-branes, or current algebras on the celestial sphere \cite{Strominger:2013lka,Strominger:2013jfa,Costello:2022wso,Guevara:2021abz,Strominger:2021mtt,Freidel:2021hws,Kmec:2025ftx}. 

In a quantum theory, these classical algebras can, in general, be deformed or explicitly broken by Wilsonian or loop corrections; see, e.g., \cite{Elvang:2017,
Mago:2021wje,Bhardwaj:2022anh,Bittleston:2022jeq,Costello:2022aol}. 
In QCD, for example, quantum confinement eliminates the soft modes and the algebra entirely. Some examples of full quantum theories with soft algebras are known.
A large class of twistorial quantum theories in flat space, with soft algebras related to quantum integrability, were discovered in \cite{Costello:2022wso,Costello:2022jpg,Costello:2023hmi,Kmec:2025ftx}.\footnote{These theories have a trivial $\cal S$-matrix, but $S$ and ${\cal L}w_{1+\infty}$ act on form factors.} Recently it was shown \cite{Sheta:2025oep,Wei:2026cft3} that, for unconfined bulk gauge theories in AdS$_4$/CFT$_3$, the 
$S$-algebra remains undeformed in the full quantum theory, while the cosmological constant deforms ${\cal L}w_{1+\infty}$ to ${\cal L}_{\Lambda}w_{1+\infty}$ \cite{Taylor:2023ajd,Bittleston:2024rqe,Schwarz:2023celestial}. 

For nonabelian gauge theory, a nontrivial $S$-algebra as a flat space asymptotic symmetry algebra can at best exist for a theory with a conformal interacting nonabelian Coulomb phase in the deep IR; otherwise the theory either confines or is trivial.  
Even in such conformal examples, a serious obstacle is encountered: due to IR divergences, there is no known construction of an $\cal S$-matrix on which the $S$-algebra may act. Standard regularization procedures induce deformations of the soft theorems beyond leading order, and logarithmic terms appear at loop level \cite{Bern:2014oka,Bonocore:2015esa,Brandhuber:2015vhm,DelDuca:1990gz}.  The color structure in nonabelian gauge theories introduces an additional complication \cite{Magnea:2021fvy,Gonzalez:2021dxw,Magnea:2025zut}, since gluons can interact with each other. 
Accordingly, it has been speculated that an uncorrected quantum-exact $S$-algebra can exist only in twistorial-type theories with no IR divergences \cite{Kmec:2025ftx,Bittleston:2022jeq,Fernandez:2023kdu}. This paper considers the existence of an $S$-algebra for standard 4D gauge theories including quantum corrections.

Planar ${\cal N}=4$ SYM offers a unique arena to address these questions. It is UV finite, maximally supersymmetric, and exactly conformal.\footnote{By this we mean dilation/Lorentz invariant: special conformal transformations have an anomaly in flat space \cite{Beisert:2010gn,Larkoski:2014hta}.} Its color-ordered amplitudes are understood to an amazing level of detail, including BDS exponentiation of infrared divergences, universal collinear factorization, dual conformal symmetry and the amplitude/Wilson-loop duality \cite{Anastasiou:2003kj,Bern:2005iz,Alday:2007hr,Drummond:2007aua,Brandhuber:2007yx,Drummond:2007au,Arkani-Hamed:2010zjl,Arkani-Hamed:2022rwr}. 

In this paper we exploit this detailed knowledge to provide evidence for the existence of an undeformed $S$-algebra action in planar ${\cal N}=4$ SYM. We argue that color-ordered gluon amplitudes can be written in a judiciously chosen factorized form $\mathcal{A}_n \;=\; \mathcal{A}^{\rm soft}_n \times \mathcal{A}^{\rm hard}_n$, where $\mathcal{A}^{\rm soft}_n$ is IR divergent -- but kinematically one-loop exact -- while $\mathcal{A}^{\rm hard}_n$ is IR-finite and encodes the higher-loop corrections contained in the remainder function together with the finite ratio function.\footnote{We note that, in contrast to some other discussions, the tree-level collinear divergences are included in $\mathcal{A}^{\rm hard}_n$, while the loop-level collinear divergences, characterised by the collinear anomalous dimension, are included in $\mathcal{A}^{\rm soft}_n$.}  We further argue that $\mathcal{A}^{\rm hard}_n$ provides a representation of the undeformed tree-level $S$-algebra. It also obeys both the uncorrected tree-level leading soft theorem and tree-level splitting.\footnote{At the formal level of the {\it integrand}, the leading soft theorem was studied in \cite{Bianchi:2014gla}. The aim of the present paper is to study soft theorems at the level of finite {\it integrals}, and to set up a computational framework to go beyond leading order.} { Corrections to subleading soft theorems are not precluded, but they are subject to integrability conditions by the existence of an undeformed $S$-algebra.}

Our argument rests on several stated assumptions about detailed properties of the dimensionally regulated scattering amplitudes. All of them are implicit or explicit in various places in the literature. They concern the structure of IR divergences, collinear factorization properties of scattering amplitudes, the one-loop exponentiation of the splitting function, and the duality with polygonal null Wilson loops \cite{Bern:2005iz,Bern:1994zx,Brandhuber:2007yx,Drummond:2007au,Basso:2013vsa,Basso:2013aha,Basso:2014koa,Arkani-Hamed:2010zjl}. This last assumption, in particular, leads to a way to compute corrections around the soft, collinear, and half-collinear limits, and to obtain leading universal logarithmic pieces to arbitrary order in perturbation theory. 

A crucial role is played by the nonrenormalization of the leading pole in the half-collinear $p_1\cdot p_2 \to 0$ splitting functions for $\mathcal{A}^{\rm hard}_n$, which give the structure constants of the soft algebra. In the loop expansion, the corrections are proportional to powers of $\log(p_1\cdot p_2)$. These diverge and naively seem to correct the half-collinear limits appearing in the soft algebra. However, using results of \cite{Basso:2013vsa,Basso:2013aha,Basso:2014koa,Basso:2014pla,Drummond:2015jea}, we argue in section \ref{sec:collinearnp} that they essentially resum to 
$(p_1\cdot p_2)^{c}$ with $c$ a nonnegative function of the coupling $g_{YM}$. Hence they do not correct the leading pole or the $S$-algebra at finite coupling. 

The tower of subleading soft theorems and the $S$-algebra in some contexts (including tree-level YM, see section 5) follows  directly  from the leading soft theorem using $SO(4,2)$ conformal invariance \cite{Sheta:2025oep} (the first subleading theorem is so derived in \cite{Larkoski:2014hta}). Thus the nonrenormalization of the leading soft theorem already suggests an $S$-algebra for planar ${\cal N}=4$. However, the derivation of \cite{Sheta:2025oep} relies crucially on special conformal symmetry 
which is anomalous for ${\cal N}=4$ \cite{Beisert:2010gn}. This necessitated the direct analysis of the amplitudes presented here which arrives at the same conclusion. 

A similar factorization of the amplitudes for both QED and gravity in the form $\mathcal{A}_n \;=\; \mathcal{A}^{\rm soft}_n \times \mathcal{A}^{\rm hard}_n$ was studied in the celestial basis in \cite{Arkani-Hamed:2020gyp}. In those contexts, $\mathcal{A}^{\rm hard}_n$ was shown to be the actual IR-finite physical scattering amplitude of the generalization of Faddeev-Kulish dressed states introduced in \cite{Kapec:2017tkm,Choi:2017bna,Choi:2017ylo}. The construction of such an IR-finite basis of scattering states for nonabelian gauge theory is an important unsolved problem. IR divergences are intimately related to soft symmetries \cite{Kapec:2017tkm,Choi:2017bna,Choi:2017ylo,Arkani-Hamed:2020gyp,Nande:2017dba}. We hope to extend the interpretation of \cite{Arkani-Hamed:2020gyp} to nonabelian gauge theory and realize $\mathcal{A}^{\rm hard}_n$ as the physical scattering amplitude of appropriately soft-gluon-dressed states.

This paper is organized as follows. Section 2 treats MHV amplitudes. We factor out the Parke-Taylor tree amplitude, write the logarithm of the remaining function as a universal infrared-divergent term plus a finite term $F_n$, and use the Wilson-loop duality and the anomalous dual conformal Ward identity to decompose $F_n$ into the BDS one-loop solution and the dual-conformal remainder $R_n$. The all-loop splitting function carries the loop-dependent collinear singularity, while the finite remainder has a smooth collinear limit for a judicious choice of a function $h_n(a)$ appearing in the definition of $R_n$. The actual soft limit is approached as an endpoint of a collinear limit. We explain why it is still smooth for the remainder: the Wilson-loop OPE organizes the approach to the limit by flux-tube states, and in the Euclidean region the relevant Fourier integrals do not create new endpoint growth for finite coupling. The section also checks the statement in the two-loop hexagon remainder and discusses continuation to the physical Mandelstam region. Along the way, the assumptions on which our argument relies are explicitly stated. At the end of the section, in \ref{sec:collinearnp}, we show that the half-collinear limit is also smooth, which is relevant for the $S$-algebra. 

Section 3 extends the story beyond MHV. A general superamplitude is written as the MHV superamplitude times the finite ratio function ${\cal P}_n$, so the infrared divergences stay in the MHV factor. We study the soft limit in momentum twistor variables, where the degeneration of the soft leg is especially transparent. At tree level, the NMHV ratio function reduces by standard identities among $R$-invariants. At one loop, the explicit six-point example and the general $n$-point form reduce in the same way. We also use the supersymmetric Wilson-loop OPE to show that the half-collinear limit of ${\cal P}_n$ is uncorrected from its tree value. The result is that ${\cal P}_n\to{\cal P}_{n-1}$ smoothly in both limits, giving the non-MHV input needed for a hard amplitude with the ordinary tree-level leading soft behavior and the tree-level half-collinear pole. 

Section 4 combines the MHV and non-MHV inputs into the soft/hard factorization used in the rest of the paper,
$\mathcal A^{\rm soft}_n(\epsilon)=e^{G_n(\epsilon)}$ and
$\mathcal A^{\rm hard}_n=\mathcal A^{\rm tree}_{n,\rm MHV}e^{R_n}{\cal P}_n$.
With this choice all infrared poles and the BDS-like soft data are assigned to
$\mathcal A^{\rm soft}_n$, while $\mathcal A^{\rm hard}_n$ is IR finite, encodes the higher loop corrections to scattering, has the tree-level adjacent
collinear and half-collinear poles and obeys the
uncorrected tree-level leading soft theorem. The same definition also removes the dual conformal anomaly
from the hard factor, rendering $\mathcal A^{\rm hard}_n$ dual-conformally
covariant.

Section 5 applies these properties to show that the hard amplitudes furnish a representation of the undeformed $S$-algebra. At loop level, the 
tower of $S$-algebra generators is awkward to define in an $\omega$ basis of energy eigenstates because of $\log \omega$ ambiguities in the definitions. The analysis is simpler in a celestial basis of Lorentz/conformal primary scattering states, in which the soft generators are the residues at boost weights $\Delta=1,0,-1,\ldots$
of the positive-helicity gluon operators, denoted $\mathsf S^{p,a}_{\bar m,m}$ with
$p =\frac{3-\Delta}{2}=1,\frac{3}{2},2,\ldots$.  The leading holomorphic gluon OPE, as defined from $\mathcal A^{\rm hard}$, determines the algebra of soft insertions as  the
Mellin transform of the uncorrected half-collinear pole. The soft theorems then equate $\mathcal A^{\rm hard}_{n+1}$ with a soft gluon insertion $\mathsf S^{p,a}_{\bar m,m}$ to a differential operator
$\mathsf T^{p,a}_{\bar m,m}$ on $\mathcal A^{\rm hard}_n$, whose tree level expression is given explicitly.  Because the insertion algebra
itself is fixed by the tree-level hard splitting function, any quantum 
corrections to $\mathsf T^{p,a}_{\bar m,m}$ must satisfy the corresponding undeformed $S$-algebra
integrability conditions.

Section 6 discusses a few open questions. The appendix develops the
multi-channel Wilson-loop OPE calculation of the leading nontrivial soft
correction, including the universal
$\delta\,a^\ell\log^\ell\delta$ term and its geometric cross-ratio map.

\section{Leading Soft theorems for MHV amplitudes}

\subsection{Collinear behaviour of MHV amplitudes}

Let us denote by ${A}_n^{MHV}$ the color-ordered MHV $n$-gluon amplitude in planar ${\cal N}=4$ SYM. The amplitude depends on both the helicities and the momenta of the external gluons. The helicity dependence is fully encoded in the tree-level amplitude, given by the Parke-Taylor formula
\begin{equation}
A_{n,tree}^{MHV}(1^+ \cdots i^-, j^- \cdots n^+)=i \frac{\langle i j \rangle^4}{\langle 1 2 \rangle \langle 2 3 \rangle \cdots \langle n 1 \rangle}.
\end{equation}
Factoring out the tree-level amplitude, we define the normalized amplitude $M_n(p_1,\cdots,p_n)$
\begin{equation}
\label{rationM}
A_n^{MHV} = A_{n,tree}^{MHV} \times M_n,
\end{equation}
The normalized amplitude depends only on the external momenta, which are on shell, $p_i^2=0$. We will suppress the dependence on the external helicities and momenta whenever they are clear. At loop order, the amplitude has IR singularities, which we regularise using dimensional regularisation in $d=4-2\epsilon$. These singularities have the universal exponential form \cite{Magnea:1990zb,Sterman:2002qn,Bern:2005iz}
\begin{equation}
\label{IRdivs}
\log M_n = -\frac{1}{4} \sum_{\ell=1}^\infty a^\ell \left( \frac{f^{(\ell)}}{(\ell \epsilon)^2} + \frac{g^{(\ell)}}{(\ell \epsilon)}   \right) \sum_{i=1}^n \left( \frac{-s_{i,i+1}}{\mu_{IR}^2}\right)^{-\ell \epsilon} + F_n(p_1,\cdots,p_n) +{\cal O}(\epsilon)
\end{equation}
where $a=\frac{g_{YM}^2 N}{8\pi^2}$ is the coupling constant, $\mu_{IR}$ is the IR cut-off, $s_{i,i+1}=(p_i+p_{i+1})^2=2 p_i \cdot p_{i+1}$, and $f(a) = \sum_{\ell=1}^\infty a^\ell f^{(\ell)}, ~~g(a) = \sum_{\ell=1}^\infty a^\ell g^{(\ell)}$ are two functions of the coupling constant $a$, commonly denoted the cusp and collinear anomalous dimensions. At weak coupling, in this normalization, $f(a)=2a+\cdots$. $F_n(p_1,\cdots,p_n)$ is a finite contribution, also a function of the coupling constant. We now make the following assumption:

\bigskip

\noindent {\bf Assumption 1}: Scattering amplitudes in planar MSYM are dual to polygonal null Wilson loops. 

\bigskip

\noindent This duality was first observed at strong coupling \cite{Alday:2007hr} and soon after at weak coupling in a variety of examples, involving MHV \cite{Drummond:2007aua,
Brandhuber:2007yx,Drummond:2007cf,Drummond:2007au,
Drummond:2008aq,Bern:2008ap} and non-MHV \cite{Mason:2010yk,Caron-Huot:2010ryg} amplitudes. In \cite{Mason:2010yk}, the duality was also shown to hold formally at the level of the integrand. It maps the normalized amplitude $M_n(p_1,\cdots,p_n)$ to the expectation value of a Wilson loop ${\cal W}_n(p_1,\cdots,p_n)$ for a polygonal null contour whose edges are the momenta $p_i$. We will rely on several aspects of this duality. Here we use the fact that it implies dual conformal symmetry \cite{Drummond:2006rz,Drummond:2007au,Drummond:2008vq,Berkovits:2008ic}. As a consequence, the function $F_n(p_1,\cdots,p_n;a)$ satisfies a Ward identity \cite{Drummond:2007au}. This Ward identity is anomalous due to IR divergences or, rather, the UV divergences of the dual Wilson loop. More precisely, if we define null-separated points $x_i$ such that 
\begin{equation}
p_i = x_{i} - x_{i+1},~~~~x_{n+i} \equiv x_i,
\end{equation}
then $F_n(p_1,\cdots,p_n) \to F_n(x_1,\cdots,x_n)$, written in terms of these dual variables, is invariant under rotations, translations, and dilatations acting on the $x_i$. On the other hand, dual special conformal transformations are anomalous. The anomaly depends on the coupling only through its normalization, given by the cusp anomalous dimensions, and the anomalous Ward identity takes the form
\begin{equation}
K^\mu F_n(x_1,\cdots,x_n)= \frac{1}{2} f(a) \sum_{i=1}^n(x^\mu_i-x^\mu_{i+1}) \log \left( \frac{x_{i,i+2}^2}{x_{i-1,i+1}^2}\right),
\label{dualKan}
\end{equation}
where 
\begin{equation}
K_\mu = \sum_{i=1}^n \left(2 x_{i \mu} x_i \cdot \frac{\partial}{\partial x_i} -  x_i^2 \frac{\partial}{\partial x_i^\mu} \right).
\end{equation}
are the generators of the dual special conformal transformations. A particularly natural solution to (\ref{dualKan}) is given by the one-loop answer $F^{(1)}_n(x_1,\cdots,x_n)$ multiplied by $\frac{1}{2} f(a)$, to give the correct coupling-constant dependence. This corresponds to the BDS ansatz \cite{Bern:2005iz}. In addition, we can always add a term annihilated by all dual conformal transformations. The general answer is then given by
\begin{equation}
\label{WIsol}
F_n(x_1,\cdots,x_n;a) = \frac{1}{2} f(a) F^{(1)}_n(x_1,\cdots,x_n) + R_n(u_1,u_2,\cdots)+h_n(a),
\end{equation}
where the {\it remainder function} $R_n(u_1,u_2,\cdots)$ is an arbitrary function of dual-invariant cross-ratios
\begin{equation}
u_{ijkl}=\frac{x_{ij}^2 x_{kl}^2}{x_{ik}^2 x_{jl}^2}.
\label{cr}
\end{equation}
Such invariants exist starting from $n=6$. For the scattering of $n$ gluons, there are $3(n-5)$ independent cross-ratios. The constant function $h_n(a)$, whose dependence on $n$ will be fixed momentarily, is added to ensure $R_n \to R_{n-1}$ smoothly in the collinear limit. This is a well-established behaviour\footnote{For example, the same conclusion was reached in \cite{Bern:2008ap} following similar arguments.} and is inherent in the Wilson-loop OPE to be introduced below, but for our purposes it will be important to review the assumptions on which it relies. Note that, by construction, the remainder function and $h_n(a)$ start at two loops. 

Let us now analyse the collinear limit of the MHV amplitudes, where two consecutive legs $p_j,p_{j+1}$ become collinear:
$$p^2 = (p_j+p_{j+1})^2 \to 0,~~~p_j \approx z p,~~p_{j+1} \approx (1-z) p$$
This limit is governed by the well-studied splitting functions. For MHV amplitudes, ${\cal N}=4$ Ward identities imply that the full helicity structure is contained in the tree-level splitting function \cite{Bern:1994zx}. Factoring out the tree-level amplitude, we can then define a splitting function which depends only on $z$, $\epsilon$, and $s=p^2$. We then make the following assumption \cite{Bern:1994zx,Anastasiou:2003kj,Bern:2004cz,Bern:2005iz}

\bigskip

\noindent {\bf Assumption 2}: The behavior of the normalized amplitude $M_n$ in the collinear limit is given by 
\begin{equation}
M_n(\cdots, p_j, p_{j+1}, \cdots) \to r(\epsilon,z,s_{j,j+1}) M_{n-1}(\cdots, p, \cdots),
\end{equation}
where $r(\epsilon,z,s_{j,j+1})$ is the splitting amplitude normalized by the splitting amplitude at tree level. Note that the splitting function does not depend on $n$ or on non-adjacent momenta. At one loop, it is given by
\begin{equation}
r^{(1)}(\epsilon,z,s) = \frac{\hat c_\Gamma}{\epsilon^2} \left( \frac{\mu_{IR}^2}{-s} \right)^{\epsilon} \left( -\frac{\pi \epsilon}{\sin(\pi \epsilon)} \left( \frac{1-z}{z} \right)^{\epsilon} + 2 \epsilon \log(1-z) + {\cal O}(\epsilon^3)\right),
\end{equation}
with 
\begin{equation}
\hat c_\Gamma = \frac{e^{\epsilon \gamma_{\rm E}}}{2} \frac{\Gamma(1+\epsilon)\Gamma^2(1-\epsilon)}{\Gamma(1-2\epsilon)},
\end{equation}
where $\gamma_{\rm E}$ is Euler's constant. At two loops, the splitting function was computed in \cite{Anastasiou:2003kj}, where a recursion relation was noted. This led to the following assumption \cite{Bern:2005iz}

\bigskip

\noindent {\bf Assumption 3}: At higher loops, the splitting function exponentiates.
\begin{equation}
r(\epsilon,z,s) = \exp \left[ \frac{1}{2} \sum_{\ell=1}^\infty a^\ell \left(f^{(\ell)} +  \ell \epsilon\, g^{(\ell)} \right)r^{(1)}(\ell \epsilon,z,s) + \hat r(a) + {\cal O}(\epsilon) \right],
\end{equation}
with $\hat r(a)$ a function of the coupling constant which starts at two loops. This is a very strong assumption. Combined with the explicit one-loop results $F_n^{(1)}(p_1,\cdots,p_n)$ in \cite{Bern:2005iz}, it implies
\begin{eqnarray}
\label{collinearBDS}
&& F_n^{(1)}(\cdots, p_j, p_{j+1},\cdots ) \xrightarrow{coll} F_{n-1}^{(1)}(\cdots, p,\cdots ) 
\\
&&~~~~~~~~ -\frac{1}{2} \left( \log z \log \left(\frac{ p^2}{2p_{j-1} \cdot p} \right)+\log(1-z) \log \left(\frac{p^2}{2zp_{j+2} \cdot p} \right)   \right)- \frac{\pi^2}{24} \nonumber
\end{eqnarray}
This can be combined with the corresponding limit in the IR-divergent part of $\log M_n$ in (\ref{IRdivs}), where we take
\begin{equation}
s_{j-1,j} \to 2 z p_{j-1}\cdot p,~~~s_{j+1,j+2} \to 2 (1-z) p_{j+2}\cdot p,~~~s_{j,j+1}=p^2,
\end{equation}
with $p^2 \to 0$. The two contributions combine exactly into the splitting function, and consistency with assumptions 2 and 3 requires
\begin{equation}
R_n(\cdots, p_j,p_{j+1},\cdots) +h_n(a) \xrightarrow{coll} R_{n-1}(\cdots,p,\cdots)+\hat r(a) +h_{n-1}(a).
\end{equation}
We then make the choice $h_n(a) = n \hat r(a) + h_0(a)$, fixing the $n$-dependence of the function in $(\ref{WIsol})$, and fix the constant $h_0(a)$ so that the remainder function for four points is exactly zero. Then universality of the collinear limit implies that the remainder function is also zero for five points. With this choice, it follows that
\begin{equation}
\boxed{\displaystyle R_n(\cdots, p_j,p_{j+1},\cdots) \xrightarrow{coll} R_{n-1}(\cdots,p,\cdots)}
\end{equation}
smoothly, without additive constants. 

\subsection{Soft from collinear}
\label{sect:softfromcol}

Having determined the collinear limit of the remainder function, we would like to determine its soft limit. Formally, the soft limit is a particular case of the collinear limit. However, it corresponds to a degenerate limit with $z \sim s \to 0$ in the language of the splitting function. When discussing the splitting properties of scattering amplitudes, one assumes $s \to 0$ while keeping $z$ finite. Terms invisible in the collinear limit may contribute when $z \sim s \to 0$. Consider, for definiteness, the six-point remainder function $R_6(u_1,u_2,u_3)$, which depends on three independent cross-ratios
\begin{equation}
\label{hexagonratios}
u_1 = \frac{x^2_{13} x_{46}^2}{x_{14}^2 x_{36}^2},~~u_2 = \frac{x^2_{24} x^2_{15}}{x^2_{25} x^2_{14}},~~u_3 = \frac{x^2_{35} x^2_{26}}{x^2_{36} x^2_{25}}.
\end{equation}
We always approach the collinear and soft limits from the Euclidean region; namely, all separations $x_{ij}$ are taken to be space-like unless $i$ and $j$ are consecutive, in which case the separation is null. In terms of dual coordinates, a soft limit corresponds to two points colliding. For example, consider $x_6 = x_1 +\delta q$ with $\delta \to 0$. In terms of cross-ratios, the soft limit corresponds to
\begin{equation}
\text{Soft:} ~~~~~u_1 =1+ {\cal O}(\delta),~~~u_2 = {\cal O}(\delta),~~~u_3 = {\cal O}(\delta)
\end{equation}
Consider now the collinear limit. In terms of dual coordinates, we take $x_{15}^2 \to 0$, with $x_6$ approaching the line $x_{15}$. The results in the previous section for the collinear limit of the remainder function imply (we will be more precise below)
\begin{equation}
\text{Collinear:} ~~~~~  \lim_{\eta \to 0} R_6(1-u, \eta,u) = 0,~~~0<u<1.
\end{equation}
We can then reach the soft limit through the collinear limit by further taking $u \to 0$. This, of course, is a subtle limit, as subleading corrections to the collinear limit may be enhanced as we take $u \sim \eta \to 0$ \footnote{Consider, for instance, a term of the form $\log(1-u_2/u_3)$. Such a term would be invisible in the leading collinear limit, but would contribute in the soft limit. Of course, such a term is forbidden since it has a branch point at $u_2=u_3$. For a physical massless scattering amplitude, branch points can occur only when $x_{i,j}^2 =0$ for non-consecutive $i$ and $j$. In terms of the cross-ratios, the branch points in the Euclidean region can then be located only at $u_i = 0$ or $u_i=\infty$ for some $u_i$; see the discussion in \cite{Dixon:2020cnr}.}. We will show below that this is not the case by considering the full tower of collinear corrections and showing that there is no enhancement as we reach the soft limit. The punchline is that, at leading order in the $\delta \to 0$ soft limit, the remainder function satisfies
\begin{equation}
\boxed{\displaystyle R_n(\cdots, p_{j-1}, \delta q,p_{j+1},\cdots) \xrightarrow{soft} R_{n-1}(\cdots,p_{j-1}, p_{j+1},\cdots),}
\end{equation}
again without any additive constants. Here $q^2=0$ and the momenta follow a smooth on-shell, momentum-conserving soft path $p^{(\delta)}_{j \pm 1} \to p_{j \pm 1}$. One explicit path is given in the appendix.

\subsubsection{Collinear expansions}
\begin{figure}[h]
\centering
\includegraphics[width=1\textwidth]{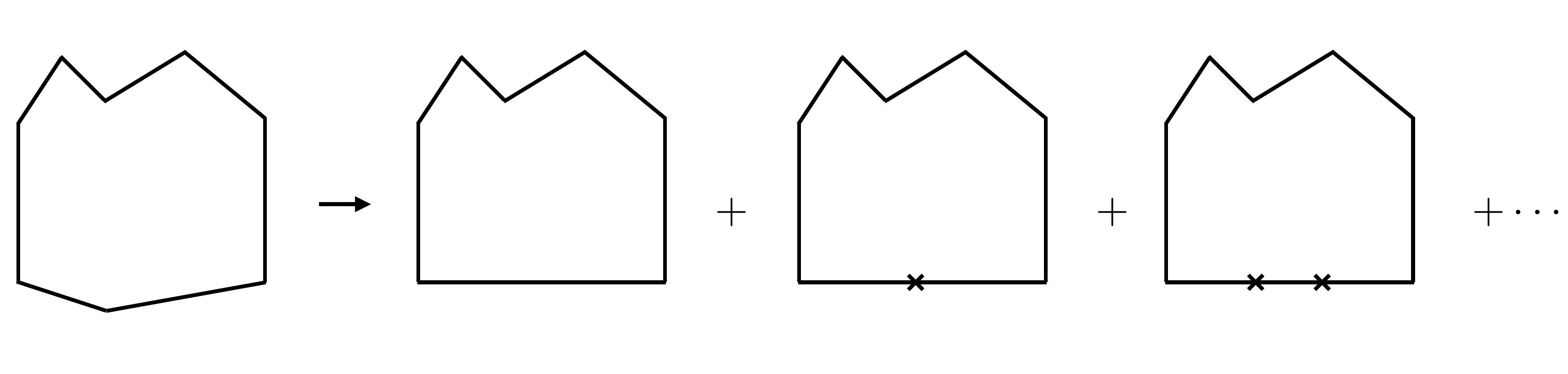}
\caption{Collinear limit of a polygonal null Wilson loop. Subleading corrections correspond to the insertion of operators along the null line at the bottom.}
\label{fig1}
\end{figure}

The duality between scattering amplitudes in planar MSYM and polygonal null Wilson loops -- Assumption 1 -- has another remarkable consequence: it makes it possible to control, and in many cases exactly compute, subleading corrections to the collinear limit. Expansions around the collinear limit of polygonal null Wilson loops were systematically studied in \cite{Alday:2010ku}, where the notion of an operator product expansion for Wilson loops was introduced. This was developed much further in \cite{Gaiotto:2011dt,Caron-Huot:2011dec,Basso:2013vsa,Basso:2013aha,Basso:2014koa}, where it was converted into a powerful computational tool. For our purposes, we shall consider the simplest case, where two consecutive lines become collinear; see figure \ref{fig1}.
At leading order, we recover a polygonal null Wilson loop with one less side. The first correction corresponds to the insertion of one operator along the contour that has become collinear, the second to the insertion of two operators, and so on. 

\begin{figure}[h]
\centering
\includegraphics[width=0.9\textwidth]{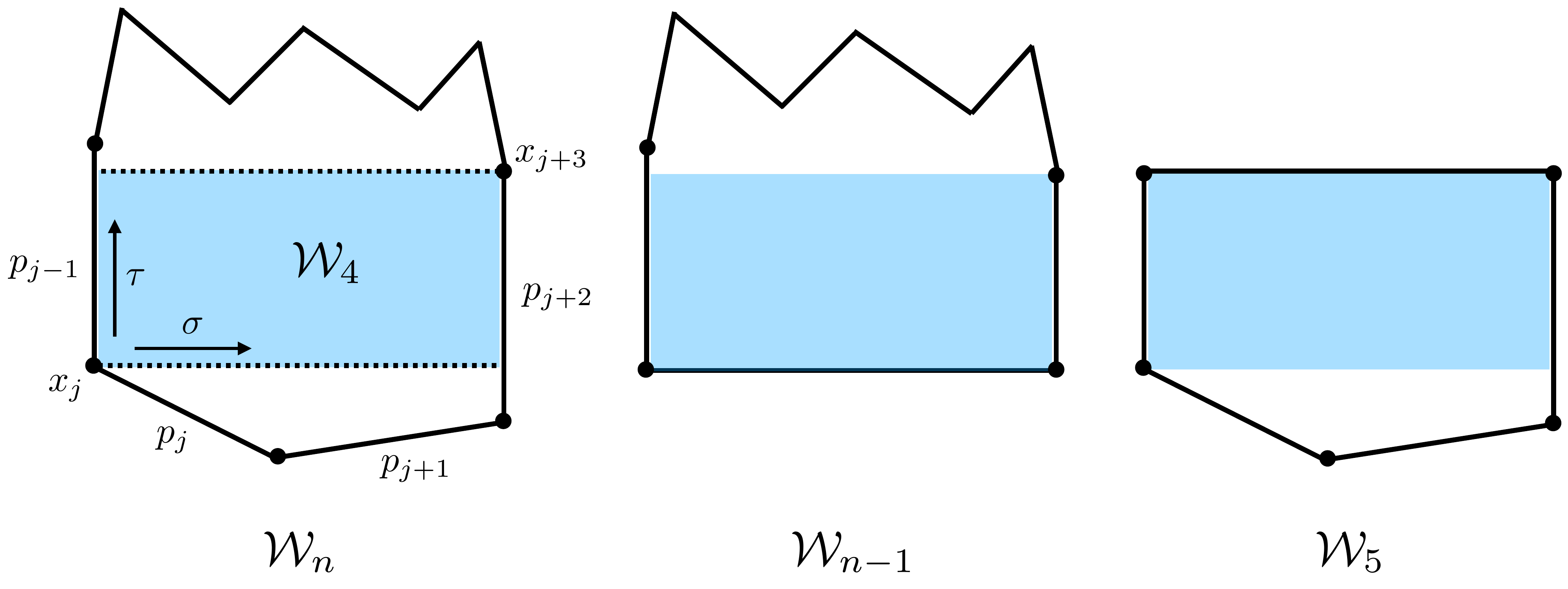}
\caption{The reference square splits the original polygon into a top polygon ${\cal W}_{n-1}$ and a bottom polygon ${\cal W}_{5}$. We also show two of the symmetries of the reference square.}
\label{fig2}
\end{figure}

More precisely, take a polygonal null Wilson loop with $n$ sides ${\cal W}_n$, whose sides are the null momenta $p_i$, and consider the limit where the null momenta $p_j$ and $p_{j+1}$ become collinear; see figure \ref{fig2}. 
We then construct a reference null Wilson loop with four sides ${\cal W}_4$ such that two opposite null sides are parallel to $p_{j-1}$ and $p_{j+2}$, and the other two sides are also null and contain $x_j$ and $x_{j+3}$; see figure \ref{fig2}. This reference square loop splits the original polygon into a top polygon ${\cal W}_{n-1}$ and a bottom polygon ${\cal W}_5$. We can then consider the following ratio
\begin{equation}
{\cal R}_n = \log \left( \frac{{\cal W}_n {\cal W}_4}{{\cal W}_{n-1} {\cal W}_5} \right)
\end{equation}
which is finite as a consequence of the universal structure of UV divergences for polygonal null Wilson loops -- which mimics that of IR divergences in scattering amplitudes -- and (dual) conformally invariant. We will draw conclusions for this ratio function and then translate them to the remainder function $R_n$. The two are simply related. 

The idea behind developing an OPE/collinear expansion for ${\cal R}_n$ is the following. The reference square is invariant under three commuting symmetries\footnote{This can be understood as follows. By a conformal transformation, map the reference square to $R^{1,1}$. Then map one cusp to the origin, the opposite cusp to spatial infinity, and the other two cusps to points at null infinity. The Wilson loop is then given by the two null lines $x^+=0$ and $x^-=0$ with $x^+,x^->0$. The three symmetries are then rescalings of $x^+$, rescalings of $x^-$, and rotations in the transverse plane.} $R_\tau \times R_\sigma \times SO(2)_\phi$ with conjugate variables $\tau,\sigma,\phi$. The symmetry conjugate to $\tau$ has the interpretation of a Hamiltonian that propagates states from the bottom to the top of the reference square. The symmetry conjugate to $\sigma$ has the interpretation of momentum in the perpendicular direction, while $\phi$ corresponds to rotations in the transverse plane. We then act with these symmetries on the three points at the bottom $x_j,x_{j+1},x_{j+2}$. This results in a family of Wilson loops with ratio ${\cal R}_n(\tau,\sigma,\phi)$. For the case of the hexagon with $n=6$, the three standard cross-ratios $u_1,u_2,u_3$ can be written in terms of $\tau,\sigma,\phi$, and we give such a parametrization below. For $n > 6$, the ratio ${\cal R}_n$ is a function of $\tau,\sigma,\phi$ plus $3(n-6)$ other independent cross-ratios, which we denote collectively as ${\bf v}$. The function ${\cal R}_n(\tau,\sigma,\phi,{\bf v})$ then admits an expansion in terms of eigenstates of the symmetries $R_\tau,R_\sigma$, and $SO(2)_\phi$, propagating along the reference square from the bottom to the top
\begin{equation}
{\cal R}_n(\tau,\sigma,\phi,{\bf v}) = \sum_{k,m} \int dp \, e^{i m \phi} e^{i p \sigma -E_k(p)\tau} \,C_{k,m}(p,{\bf v}),~~~C_{k,m}(p,{\bf v})=C^{bottom}_{k,m}(p)C^{top}_{k,m}(p,{\bf v}).
\label{WLOPE}
\end{equation}
The functions $C^{bottom}_{k,m}(p), C^{top}_{k,m}(p,{\bf v})$ are the analogues of OPE coefficients and give the overlap of the intermediate operator with the bottom/top Wilson loops. In \cite{Alday:2010ku} the following was conjectured

\bigskip

\noindent {\bf Assumption 4}: { The OPE expansion for polygonal null Wilson loops (\ref{WLOPE}) converges (uniformly) for all $\tau,\sigma,\phi$ in the Euclidean region with all non-consecutive distances space-like. }

\bigskip

\begin{figure}[h]
\centering
\includegraphics[width=0.6\textwidth]{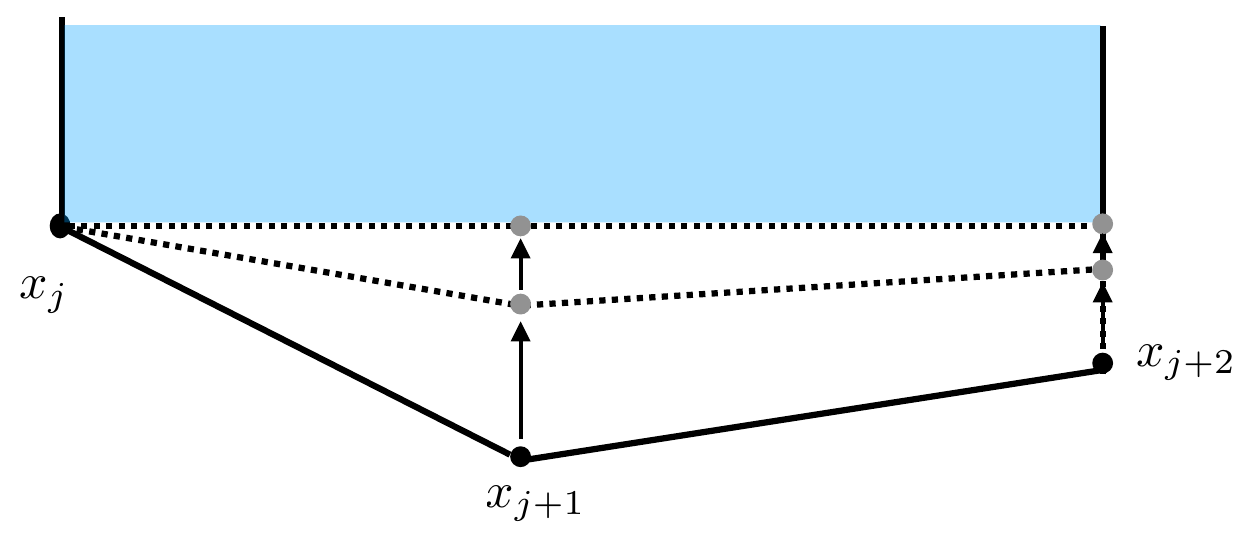}
\caption{As $\tau \to \infty$ the lines $x_{j,j+1}$ and $x_{j+1,j+2}$ approach the bottom line of the reference square, becoming collinear. If we furthermore take $\sigma \to \pm \infty$ then $x_{j+1}$ approaches either $x_j$ or $x_{j+2}$.}
\label{fig3}
\end{figure}
As shown in \cite{Alday:2010ku}, the collinear limit corresponds to $\tau \to \infty$; see figure \ref{fig3}. More precisely, as we take $\tau \to \infty$, the points $x_{j+1}$ and $x_{j+2}$ collapse into the bottom line of the reference square, and the lines $x_{j,j+1}$ and $x_{j+1,j+2}$ become collinear. The expansion around the collinear limit is then governed by the 'energy' $E_k(p)$ of the intermediate states in (\ref{WLOPE}). In perturbation theory, this is given by a positive integer plus a small correction, and we write $E_k(p)=k+\gamma_k(p)$. Furthermore, for a given $k$, only a finite set ${\cal M}_k$ of $SO(2)_\phi$ charges contributes; see \cite{Basso:2013vsa,Basso:2013aha,Basso:2014koa}. Hence, in perturbation theory\footnote{For $k>1$ there is degeneracy, but this does not modify our conclusions.}
\begin{equation}
{\cal R}_n(\tau,\sigma,\phi,{\bf v}) = \sum_{k=1}^{\infty} \sum_{m\in {\cal M}_k} e^{-k \tau} \int dp \,e^{i m \phi} e^{i p \sigma -\gamma_k(p)\tau} \,C_{k,m}(p,{\bf v}).
\label{colexp}
\end{equation}
This leads to ${\cal R}_n(\tau,\sigma,\phi,{\bf v}) \xrightarrow{\text{coll}} 0$ in the collinear limit, but it also gives a handle on arbitrarily high corrections around it. Let us now consider the soft limit. This corresponds to also taking $|\sigma| \to \infty$ with $\tau - |\sigma|$ held finite, so that $e^{|\sigma|} \sim e^\tau$. If $\sigma \to \infty$, then $x_{j+1}$ collides with $x_{j+2}$. If $\sigma \to - \infty$, then $x_{j+1}$ collides with $x_{j}$. We are then led to consider integrals of the form
\begin{equation}
f(\sigma)= \int dp \, C(p) e^{i p \sigma}
\end{equation}
and need the fact that, for the problem at hand, such integrals do not diverge exponentially, i.e. do not behave as $f(\sigma) \sim e^{|\sigma|}$ for $|\sigma| \to \infty$. Such a divergence would have the potential to enhance the factor $e^{-k \tau}$ in the soft limit. There are several ways to see that such divergences are absent in the Euclidean OPE. For instance, the integral converges at $\sigma=0$ (since this corresponds to the original Wilson loop before the action of symmetries), while for real $\sigma$ the factor $e^{ip\sigma}$ is a pure phase. More generally, the Euclidean OPE coefficients are ordinary Fourier data in the momentum $p$, and exponential growth in $\sigma$ would signal a singularity obstructing the Euclidean expansion. For example, for the simplest case of the hexagon at leading order in perturbation theory one encounters, up to an overall normalization,
\begin{equation}
C(p) = \frac{1}{1+p^2} \frac{\pi}{\cosh \left( \frac{p \pi}{2} \right)} \to f(\sigma) = \cosh \sigma \log(2 \cosh \sigma) - \sigma \sinh \sigma
\end{equation}
which actually decays exponentially for $\sigma \to \pm \infty$. Under Assumption 4, the same Euclidean OPE reasoning applies term by term, so in the collinear expansion (\ref{colexp}) of ${\cal R}_n(\tau,\sigma,\phi,{\bf v})$, we can take the soft limit without enhanced contributions. Since we are assuming the sum converges, it follows that ${\cal R}_n(\tau,\sigma,\phi,{\bf v}) \xrightarrow{\text{soft}} 0$ in the soft limit.  

We can now make the corresponding claim for the remainder function. Its relation to the ratio function ${\cal R}_n(\tau,\sigma,\phi,{\bf v})$ is given by
\begin{equation}
R_n(\tau,\sigma,\phi,{\bf v}) - R_{n-1}({\bf v})={\cal R}_n( \tau,\sigma,\phi,{\bf v}) - \frac{f(a)}{2} {\cal R}_n^{(1)}(\tau,\sigma,\phi,{\bf v}).
\end{equation}
{This follows by inserting the  decomposition $R_m=F_m-\frac{f(a)}{2}F_m^{(1)}-h_m(a)$ for each Wilson loop entering ${\cal R}_n=\log \left( {\cal W}_n {\cal W}_4/({\cal W}_{n-1}{\cal W}_5) \right)$. }
Since both terms on the right-hand side vanish in the soft limit, we have
\begin{equation}
R_n( \tau,\sigma,\phi,{\bf v})  \xrightarrow{\text{soft}} R_{n-1}({\bf v}),
\end{equation}
smoothly. 
The OPE expansion for Wilson loops can also be used to compute corrections to the soft limit and we do so in the appendix. In particular, to each order in perturbation theory $\ell$ we determine the leading correction
\begin{equation}
R^{(\ell)}_n( \tau,\sigma,\phi,{\bf v})= R^{(\ell)}_{n-1}({\bf v}) + 2 \cos \phi e^{-\tau-\sigma} P_{\rm LL}^{(\ell)}(\sigma,\tau) + \cdots
\end{equation}
for $\ell=2,3,\cdots$, with $P_{\rm LL}^{(\ell)}(\sigma,\tau)$ a homogeneous symmetric polynomial of degree $\ell$, proportional to the Narayana polynomials, and given in the appendix. Further corrections are either exponentially or power suppressed in the large $\sigma,\tau$ limit, and have not been computed.

\subsubsection{From Euclidean to Mandelstam}

\begin{figure}[h]
\centering
\includegraphics[width=0.25\textwidth]{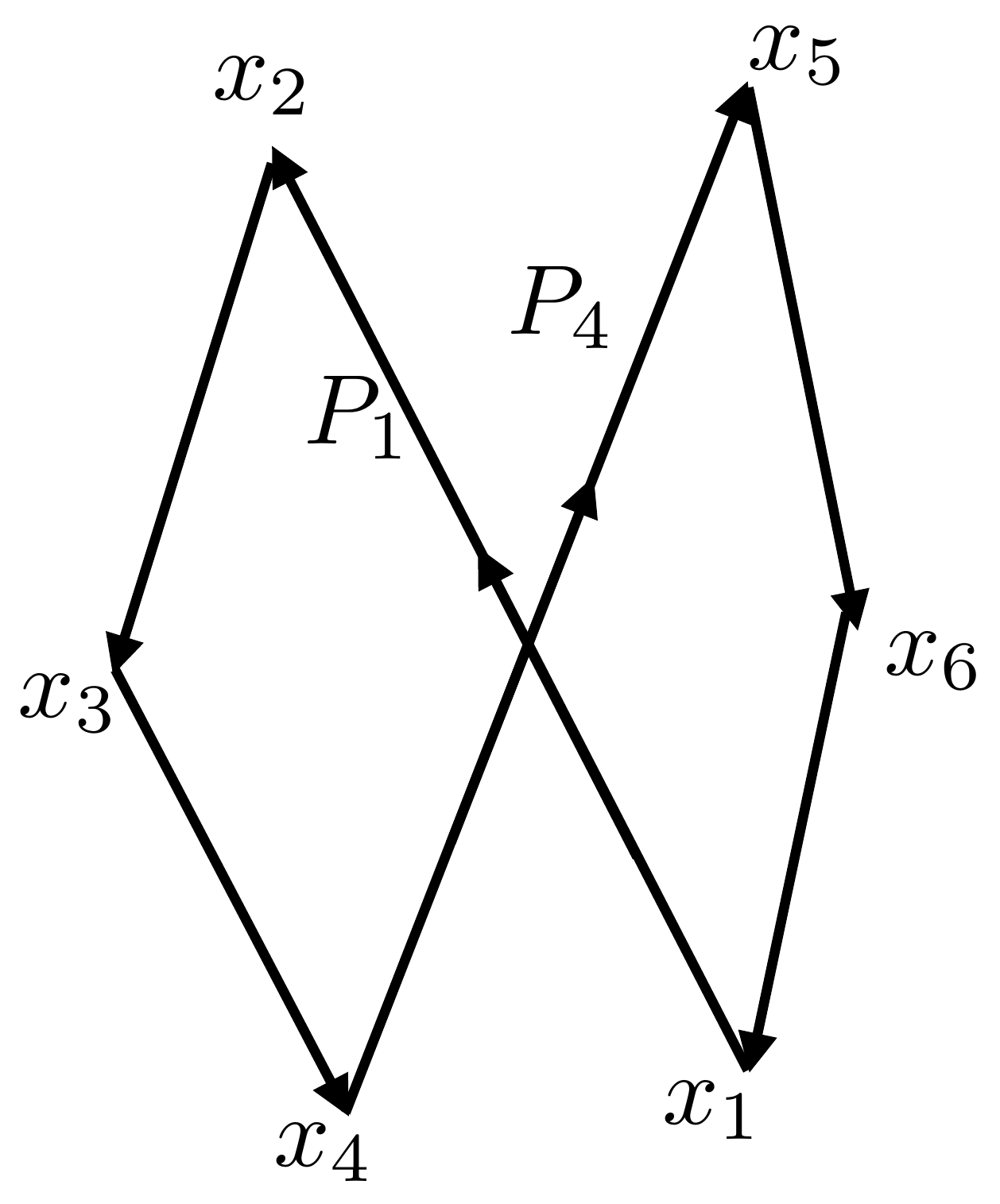}
\caption{Physical process $P_1+P_4 \to P_2+P_3+P_5+P_6$. All vectors are null, and time runs upwards.}
\label{figa4}
\end{figure}

So far our discussion has focused on the Euclidean region, where all non-consecutive distances are space-like. To go to the Mandelstam, or physical, region, we need to perform analytic continuations of the type $u_i \to e^{i \gamma \pi} u_i$. Upon such analytic continuations, subleading contributions to the collinear limit in (\ref{colexp}) can diverge exponentially for large $|\sigma|$. Indeed, this is the case for the analytic continuation relevant for Multi-Regge kinematics (MRK); see, for example, \cite{Bartels:2008ce,Hatsuda:2014oza,Basso:2014pla,Drummond:2015jea,Dixon:2016epj,Bargheer:2019lic}. The relevant analytic continuation for the soft limit, however, is different, and subleading collinear contributions do not diverge exponentially in this case. Let us see this in detail. Consider the $2 \to 4$ physical process in figure \ref{figa4}, where particles $P_1,P_4$ go into particles $P_2,P_3,P_5,P_6$, and we are in a regime such that $x_{15}$ and $x_{24}$ are time-like while all other non-consecutive distances are space-like.  
The analytic continuation from the Euclidean region to this configuration is
\begin{equation}
u_2 \to e^{-2\pi i} u_2,~~~ u_1 \to u_1,~~~ u_3 \to u_3.
\end{equation}
The physically admissible timelike collinear limits within this configuration are either $x_{24}^2 \to 0$ or $x_{15}^2 \to 0$. Both correspond to $u_2 \to 0$ with $u_1 + u_3 \to 1$. From this, we can then reach the physical soft limits $x_{23} \to 0$, $x_{24} \to 0$, etc. This corresponds precisely to the collinear expansion considered above. Furthermore, the analytic continuation corresponds essentially to taking $e^{-2\tau} \to e^{-2\pi i} e^{-2\tau}$. This can be taken term by term in (\ref{colexp}), does not affect the (uniform) convergence of the OPE and leads to the same conclusions in the soft limit. The upshot is that the analytic continuations relevant for multi-Regge kinematics and the soft limit are different, and the soft-limit statement also holds for Mandelstam or physical configurations. 

\subsubsection{Example}

The first nontrivial example corresponds to the remainder function for $n=6$ at two loops, first computed analytically in \cite{Goncharov:2010jf}. It is given by
\begin{eqnarray} 
R_6^{(2)}(u_1,u_2,u_3) &=& \sum_{i=1}^3 \left( L_4(x_i^+,x_i^-) -\frac{1}{2} Li_4(1-1/u_i)\right) \\ 
& & -\frac{1}{8} \left( \sum_{i=1}^3 Li_2(1-1/u_i) \right)^2 + \frac{1}{24} J^4+ \frac{\pi^2}{12} J^2+\frac{\pi^4}{72}, \nonumber
\end{eqnarray}
where
$$x_i^{\pm} = u_i x^{\pm},~~~x^{\pm} = \frac{u_1+u_2+u_3-1 \pm \sqrt{\Delta}}{2 u_1 u_2 u_3}$$
with $\Delta=(u_1+u_2+u_3-1)^2-4 u_1 u_2 u_3$ and we introduced the functions

\begin{equation}
L_4(x^+,x^-) = \frac{1}{8!!} \log^4(x^+ x^-) + \sum_{m=0}^3 \frac{(-1)^m}{(2m)!!} \log^m(x^+ x^-) \left( \ell_{4-m}(x^+)+ \ell_{4-m}(x^-) \right)
\end{equation}
and 
\begin{equation}
\ell_{n}(x) = \frac{1}{2} \left( Li_n(x)-(-1)^n Li_n(1/x) \right),~~ J = \sum_{i=1}^3(\ell_1(x_i^+)-\ell_1(x_i^-)).
\end{equation}
The dual conformal cross-ratios $u_k$ were given in (\ref{hexagonratios}). The expansion around the collinear limit can be described in many equivalent ways. Following \cite{Basso:2013aha}, we choose
\begin{eqnarray}
u_1 &=& \frac{e^{2 \sigma +\tau } \text{sech}(\tau )}{2 \left(2 e^{\sigma -\tau } \cos (\phi )+e^{2 \sigma }+e^{-2 \tau }+1\right)} \nonumber\\
u_2&=&\frac{1}{2} e^{-\tau } \text{sech}(\tau ), \label{hexagonparam}\\
u_3&=&\frac{1}{1+e^{2\sigma}+2 e^{\sigma-\tau}\cos \phi+ e^{-2\tau}}, \nonumber
\end{eqnarray}
In the collinear limit $\tau \to \infty$ and we have
\begin{equation}
u_1= \frac{S^2}{1+S^2},~~~u_2=0,~~~u_3=\frac{1}{1+S^2}
\end{equation}
with $S=e^\sigma$. Note that in the strict collinear limit $u_1+u_3=1$. Furthermore, the ratio $\frac{S^2}{1+S^2}$ measures one of the two longitudinal fractions in the collinear limit, depending on which daughter leg is called $z$. We find the following expansion for the remainder function around large $\tau$
\begin{equation}
R_6^{(2)}(\tau,\sigma,\phi) = \sum_{k=1}^{\infty} \sum_{m=0}^k e^{-k \tau} \cos(m \phi) \left( f^{(0)}_{k,m}(\sigma) + f^{(1)}_{k,m}(\sigma)  \tau \right)
\end{equation}
This is fully consistent with our discussion in section \ref{sect:softfromcol}. The terms linear in $\tau$ arise from the anomalous dimensions of the states propagating on the square. At loop order $\ell$, we would get terms proportional to $\tau^{0}, \tau^{1},\cdots,\tau^{\ell-1}$. One can explicitly check to any desired order that $f^{(0)}_{k,m}(\sigma), f^{(1)}_{k,m}(\sigma)$ decay exponentially for large $|\sigma|$. In general,
\begin{equation}
 f^{(0)}_{k,m}(\sigma),  f^{(1)}_{k,m}(\sigma) \sim e^{-|\sigma|},~~~ \text{for large $|\sigma|$},
\end{equation}
again in agreement with our discussion in section \ref{sect:softfromcol}. We can consider the leading correction in the soft limit. This arises from the term $k=1$ in the large-$\sigma$ limit. We obtain
\begin{equation}
R_6^{(2)}(\tau,\sigma,\phi) =e^{-\tau - \sigma} \cos \phi \left( 2 \sigma \tau+ \cdots \right) + \cdots,
\end{equation}
in perfect agreement with the results in the appendix. Finally, note that, from the point of view of section \ref{sect:softfromcol}, there is nothing too special about $n=6$ at two loops, so this is a good example of the general lessons.\footnote{In the same way, we can use four-point functions in CFT to learn about the usual OPE.}

\subsection{Collinear, half-collinear and soft limits}
\label{sec:collinearnp}

So far we have considered two limits of the remainder function, the standard collinear limit and the soft limit. In both of these limits ${\cal R}_n \to 0 $ smoothly, which implies $R_n \to R_{n-1} $ for the remainder function. In the following, we will also discuss a related half-collinear limit, which exists for complexified momenta or in signature $(2, 2)$ Klein space \cite{Arkani-Hamed:2008owk}. To describe these limits, it is convenient to parametrise the external null momenta as
\begin{equation}
p_i =\omega_i \left(1+|z_i|^2,z_i+\bar z_i,-i(z_i-\bar z_i),1-|z_i|^2 \right)
\label{ptow}
\end{equation}
such that $p_i \cdot p_j = -2 \omega_i \omega_j z_{ij} \bar z_{ij}$. Given two particles, say $1,2$, and renaming $(\omega_1,z_1,\bar z_1) \to (\omega_s,z_s,\bar z_s)$, the standard collinear and soft limits correspond to 
\begin{equation}
\text{Collinear:} ~~z_{s} \to z_2, \bar z_{s} \to \bar z_2, ~~ \omega_s ~~\text{fixed;}~~~~~~~~~\text{soft:} ~~\omega_s \to 0,~ z_s, \bar z_s~~\text{fixed},
\end{equation}
while the half-collinear limit corresponds to 
\begin{equation}
\text{Half-collinear:} ~~z_{s} \to z_2, ~~ \bar z_{s} ,\omega_s ~~\text{fixed.}
\end{equation}
The collinear and soft limits can be taken sequentially at the level of the reference hexagon cross-ratios $u_1,u_2,u_3$, and one can explicitly check that the order in which they are taken commutes. This allows us to write down the following dictionary
\begin{eqnarray}
&e^{-2\tau} \sim \omega_s z_{s2} \bar z_{s2},~~~e^{-2\sigma} \sim \omega_s, \label{dictionary}\\
&e^{-\sigma-\tau} e^{i \phi} \sim \omega_s \bar z_{s2},~~~ e^{-\sigma-\tau} e^{-i \phi} \sim \omega_s z_{s2}, \nonumber
\end{eqnarray}
where we have kept only the dependence on $\omega_s$ and $z_{s2},\bar z_{s2}$.\footnote{One can write down the complete expressions, fully consistent with $SL(2,C)$ symmetry, but they are very lengthy, and we will not need their explicit form.} From this dictionary, we can see that the usual collinear limit corresponds to taking $\tau$ large while keeping $\sigma$ fixed, whereas the soft limit corresponds to taking $\tau,\sigma$ large with $\tau-\sigma$ fixed, in full agreement with our previous discussion. In addition, the half-collinear limit corresponds to writing $i \phi \equiv \phi_e$ and then taking both $\tau,\phi_e$ large with $\tau-\phi_e,\sigma$ fixed.\footnote{In the context of the WL OPE this is known as the double scaling limit, and was originally studied in \cite{Gaiotto:2011dt}.}

We can now consider the WL OPE expansion (\ref{WLOPE}) in terms of $\omega_s,z_{s2},\bar z_{s2}$. We will do so in the regime of finite/non-zero coupling. In this regime, and for all exchanges, $E_k(p) - |m|$ is strictly positive \cite{Basso:2010in}. In addition, each term in (\ref{WLOPE}) should decay for large $\sigma$, as discussed. We can take a limit that smoothly interpolates between the collinear and soft limits in which $\tau,\sigma$ are large, with $\eta =\sigma/\tau$ fixed and $\eta \in [0,1]$. The leading behaviour for each exchange in (\ref{WLOPE}) can be computed by a saddle point and takes the form 
\begin{equation}
\int dp e^{i m \phi} e^{i p \sigma - E_k(p) \tau}\sim e^{i m \phi}  e^{-E_k(0) f(\eta) \tau} \sim \left(\frac{\bar z_{s2}}{z_{s2}} \right)^{m/2} \left( \omega_s^{1/2} |z_{s,2}|\right)^{E_k(0) f(\eta)},
\end{equation}
where $E_k(0)$ is the mass of the excitation, strictly bigger than $|m|$ at finite coupling \cite{Basso:2010in}, and $f(\eta) \geq 1, f(0)=1.$\footnote{Given that $E_k(p) =E_k(0) + \alpha p^2+\cdots$, with $\alpha>0$, for small $p$, one can explicitly check that $f(\eta)$ increases monotonically for small $\eta$. This is also true for all $\eta \in [0,1]$ in the strong coupling regime and we expect it to be true in general.} Both $E_k(0)$ and $f(\eta)$ are also functions of the coupling constant. We conclude that, for any finite, non-zero, coupling,  ${\cal R}_n(\omega_s,z_{s2},\bar z_{s2},{\bf v}) \to 0$ in all three limits, which implies $R_n \to R_{n-1}$ in terms of the remainder function. Note that this is consistent with the fact that in perturbation theory $R_n \nrightarrow R_{n-1}$ for the half-collinear limit, due to the exchange of single gluons and their bound states, for which $E_k(p) - k \sim {\cal O}(a)$:  the $z_{s2}$ logarithms are produced by the expansion in $a$.  

As an example, consider the gluon exchange at strong coupling. This was computed in \cite{Alday:2010ku}, where it was denoted $R_{\sqrt{2}}$. In this case, we obtain
\begin{equation}
R_{\sqrt{2}} = 4 \cos \phi \int {\frac{d\theta}{2\pi}} \frac{1}{(\cosh 2 \theta)^2} e^{-\tau \sqrt{2} \cosh \theta + i \sigma \sqrt{2} \sinh \theta} \sim  \cos \phi e^{-\sqrt{2} \sqrt{\tau^2+\sigma^2}}
\end{equation}
so that in this case $E(0)=\sqrt{2}$ and $f(\eta)=\sqrt{1+\eta^2}$. For the single gluon excitation, $E(0)$ increases monotonically as a function of the coupling, from one to $\sqrt{2}$.

\section{Non-MHV amplitudes}

In this section, we extend the preceding discussion to non-MHV amplitudes. It turns out that the ingredients used above also extend to this case, and hence non-MHV amplitudes are also expected to satisfy the leading soft theorem. This will be explicitly checked in various examples. 

\subsection{Argument for soft behaviour}

 Non-MHV amplitudes in ${\cal N}=4$ SYM can be conveniently described by introducing an on-shell superspace \cite{NAIR1988215,Georgiou:2004by,Arkani-Hamed:2008owk,Drummond:2008vq}. The on-shell superfield $\Phi$, depending on a Grassmann variable $\eta^A$, with $A=1,2,3,4$, is such that the positive- and negative-helicity gluons are its top and bottom components, respectively,
\begin{equation}
\Phi = G^+ +\eta^A \Gamma_A+ \cdots+ \frac{1}{4!}\eta^A \eta^B \eta^C \eta^D \epsilon_{ABCD} G^- 
\end{equation}
In on-shell superspace, all color-ordered amplitudes are combined into a single superamplitude ${\cal A}_n(\Phi_1,\cdots,\Phi_n)$. Individual helicity components can be extracted by expanding in the Grassmann variables $\eta^A_i$. The MHV superamplitude ${\cal A}_{n,MHV}$ is given by
\begin{equation}
{\cal A}_{n,MHV} = i \frac{\delta^{(4)}(\sum_i p_i ) \delta^{(8)}(\sum_i \lambda_i^\alpha \eta_i^A)}{\langle 12 \rangle \cdots \langle n 1 \rangle} M_n,
\end{equation}
where $M_n$ is exactly the normalized amplitude introduced in (\ref{rationM}) and contains all the loop information. External particles are labelled by $(\lambda_i,\tilde \lambda_i,\eta_i)$ such that their null momenta are given by $p_i^{\alpha \dot \alpha} = \lambda_i^\alpha \tilde \lambda_i^{\dot \alpha}$. The Grassmann delta function is given by
\begin{equation}
 \delta^{(8)}(\sum_i \lambda_i^\alpha \eta_i^A) = \prod_{\alpha=1}^2 \prod_{A=1}^4 \left( \sum_{i=1}^n \lambda_i^\alpha \eta_i^A \right),
\end{equation}
so that the coefficient in front of a given $(\eta_i)^4(\eta_j)^4$ reproduces the familiar $\langle i j \rangle^4$ in the numerator of the Parke-Taylor formula. We can now consider the ratio of the superamplitude to the super MHV amplitude
\begin{equation}
{\cal A}_n = {\cal A}_{n,MHV} {\cal P}_n.
\label{Pratio}
\end{equation}
The infrared divergences are entirely contained in ${\cal A}_{n,MHV}$. As a result, the ratio ${\cal P}_n$ is infrared finite. We will also assume the following, related to Assumption 1, for which \cite{Drummond:2008vq} presented strong evidence.

\bigskip

\noindent {\bf Assumption 1$^\prime$}: Scattering amplitudes in planar MSYM possess dual superconformal symmetry. 

\bigskip

While the symmetry is anomalous due to IR divergences, the anomalies in ${\cal A}_n$ are fully encoded in the MHV superamplitude. Hence ${\cal P}_n$ is dual superconformal invariant. Furthermore, as discussed in \cite{Korchemsky:2009hm}, collinear factorisation also holds for the superamplitude, with the same splitting function as for the MHV amplitude, to all loops. 

In the on-shell superspace approach, the collinear limit $(\lambda_j,\tilde \lambda_j,\eta_j) + (\lambda_{j+1},\tilde \lambda_{j+1},\eta_{j+1})  \xrightarrow{\text{coll}} (\lambda,\tilde \lambda,\eta) $ is described by \cite{Drummond:2008cr}
\begin{eqnarray}
\label{supercoll}
(\lambda_j,\tilde \lambda_j,\eta_j) &\to& ( \sqrt{z} \, \lambda, \sqrt{z}\, \tilde \lambda, \sqrt{z}\, \eta) \\
(\lambda_{j+1},\tilde \lambda_{j+1},\eta_{j+1}) &\to& ( \sqrt{1-z} \, \lambda, \sqrt{1-z}\, \tilde \lambda, \sqrt{1-z}\, \eta) \nonumber
\end{eqnarray}
so that the momenta become collinear with $p_j \to z p, p_{j+1} \to (1-z) p$ as usual, while $\lambda_j \eta_j + \lambda_{j+1} \eta_{j+1} \to \lambda \eta $. These conditions are consistent with the delta functions appearing in the superamplitude. Since we defined ${\cal P}_n$ in (\ref{Pratio}) by factoring out the MHV superamplitude, which takes care of the splitting function, we have that under the operation (\ref{supercoll})
\begin{equation}
{\cal P}_n(\cdots,\lambda_j,\tilde \lambda_j,\eta_j,\lambda_{j+1},\tilde \lambda_{j+1},\eta_{j+1},\cdots)  \xrightarrow{\text{coll}} {\cal P}_{n-1}(\cdots,\lambda,\tilde \lambda,\eta, \cdots).
\end{equation}
Next we can study corrections to this leading collinear behaviour. In the MHV case, we saw that these corrections are governed by the operator product expansion of polygonal null Wilson loops. The superamplitude can also be mapped to supersymmetrised Wilson loops \cite{Caron-Huot:2010ryg, Mason:2010yk}. Furthermore, one can develop an OPE for these superloops \cite{Sever:2011da}, and this gives the full tower of corrections around the collinear limit for superamplitudes. This has been converted into a very powerful computational tool; see, for instance, \cite{Dixon:2011nj,Dixon:2014iba}. The structure of these corrections is analogous to the structure presented in section \ref{sect:softfromcol}, but decorated by (dual) supersymmetric invariants with very simple behaviour in the soft limit. This motivates the expectation 
\begin{equation}
{\cal P}_n  \xrightarrow{\text{soft}} {\cal P}_{n-1}.
\end{equation}
We will now see this explicitly in various examples. 
\subsection{Examples}
We now discuss a variety of soft-limit examples in which we show that ${\cal P}_n\xrightarrow{\text{soft}}{\cal P}_{n-1}$.
To better understand the structure of ${\cal P}_n$, it is convenient to expand it in terms of increasing Grassmann degree:
\begin{equation}
{\cal P}_n = 1 +{\cal P}_{n,NMHV}+ {\cal P}_{n,N^2MHV} + \cdots + {\cal P}_{n,\overline{MHV}}
\end{equation}
Let us discuss ${\cal P}_{n,NMHV}$ in detail, which has Grassmann degree four. At tree level, ${\cal P}_{n,NMHV}$ is given by a linear combination of dual superconformal invariants ($R$-invariants for short). These are most easily written in terms of momentum supertwistors \cite{Hodges:2009hk,Mason:2009qx}
\begin{equation}
{\cal Z}_i = (Z_i \, | \, \chi_i),
\end{equation}
where $Z_i \in \mathbb{P}^3$ are the bosonic momentum twistors and $\chi_i$ their fermionic counterparts. They are defined by
\begin{equation}
Z_i^{\alpha,\dot \alpha} = (\lambda_i^\alpha, x_i^{\beta \dot \alpha} \lambda_{i \beta}),~~~\chi_i^A = \theta_{i}^{\alpha A} \lambda_{i \alpha},
\end{equation}
where $x_i$ and $\theta_i$ are the dual bosonic and fermionic coordinates, related to the usual momenta and $\eta_i^A$ by
\begin{equation}
p_i^{\alpha \dot \alpha}= \lambda_i^\alpha \tilde \lambda_i^{\dot \alpha}=x_{i}^{\alpha \dot \alpha}-x_{i+1}^{\alpha \dot \alpha},~~~~\lambda_i^\alpha \eta_i^A = \theta_{i}^{\alpha A}-\theta_{i+1}^{\alpha A}.
\end{equation}
Given five supertwistors ${\cal Z}_i,{\cal Z}_j,{\cal Z}_k,{\cal Z}_\ell,{\cal Z}_m$, the R-invariant 5-bracket is given by
\begin{equation}
[ijk\ell m] = \frac{\delta^{(4)}(\chi_i \langle jk\ell m \rangle + \text{cyclic} )}{\langle ijk\ell  \rangle \langle jk\ell m \rangle \langle k \ell m i \rangle \langle \ell m i j \rangle \langle m i j k \rangle},
\end{equation}
with $\langle i j k \ell \rangle=\epsilon_{ABCD} Z^A_i Z^B_j Z^C_k Z^D_\ell=\det(Z_i Z_j Z_k Z_\ell)$. The five-bracket is skew-symmetric and has weight four in the Grassmann variables. $R$-invariants satisfy linear relations. Given six momentum supertwistors labelled by $i, j, k, \ell, m, n$, we have
\begin{equation}
[ijk\ell m] - [j k\ell m n]+[k\ell m n i]-[\ell m n i j]+[m n i j k]-[i j k\ell n]=0
\end{equation}
For $n$ particles, it turns out that there are $\begin{pmatrix}
n-1 \\
4
\end{pmatrix}$ independent $R$-invariants. As we take loops into account, each $R$-invariant is multiplied by a nontrivial function of the dual conformal cross-ratios \cite{Drummond:2008vq}
\begin{equation}
u_{ijk\ell} = \frac{x_{ij}^2 x_{k\ell}^2}{x_{ik}^2 x_{j\ell}^2} =\frac{\langle i-1 \, i \, j-1 \, j\rangle \langle k-1 \, k \, \ell-1 \, \ell \rangle }{\langle i-1\, i \, k-1 \, k \rangle \langle j-1 \, j \, \ell-1 \, \ell \rangle}
\end{equation} 
which can also be written in terms of twistor four-brackets, as shown. In general, we can write
\begin{equation}
{\cal P}_{n,NMHV} = \sum_{i,j,k,\ell,m} [ijk\ell m] V_{ijk\ell m}(u_1,u_2,\cdots),
\end{equation}
where the sum runs over independent $R$-invariants. Cyclicity and parity impose some relations among them, but this is the general structure.  

To discuss the soft limit of ${\cal P}_{n,NMHV}$, we need to understand how it acts at the level of momentum twistors. This is discussed, for instance, in \cite{Bianchi:2014gla,Brandhuber:2015vhm,Dixon:2020cnr}. We would like to consider the limit in which $p_j \to 0$. In terms of dual coordinates, this means $x_j \to x_{j+1}$, which implies $x^2_{j-1,j+1} \to 0$ and $x^2_{j,j+2} \to 0$. At the level of twistors, $Z_{j-1}, Z_{j}, Z_{j+1}$ become coplanar, so that
\begin{equation}
Z_j \to c_- Z_{j-1} + c_+ Z_{j+1}.
\end{equation}
This lifts to supertwistors
\begin{equation}
{\cal Z}_j \to c_- {\cal Z}_{j-1} + c_+ {\cal Z}_{j+1}.
\end{equation}
At the level of the five brackets we take the soft limit by eliminating $j$ in favour of $j-1,j+1$. One can show
\begin{align}
&[j-1,j,j+1, a, b] \to 0 \\
&[j,j+1,a,b,c] \to \kappa_- [j-1,j+1,a,b,c] \\
&[j-1,j,a,b,c] \to \kappa_+ [j-1,j+1,a,b,c]
\end{align}
where $\kappa_++ \kappa_-=1$. Since the ratio ${\cal P}_{n,NMHV}$ is cyclically symmetric, we can choose the soft limit in which $x_n \to x_1$ and hence $Z_n \to c_- Z_{n-1} + c_+ Z_{1}$, without loss of generality. 

\subsubsection{Tree level}

The result ${\cal P}^{(0)}_{n,NMHV}$ at tree level is given by 
\begin{equation}
\label{tree-level-P}
{\cal P}^{(0)}_{n,NMHV} = \sum_{1 < i<j  < n} [1 \, i \, i+1\, j \, j+1],
\end{equation}
see, {\it e.g.}, \cite{Elvang:2009ya}. Note that the term is zero for $j=i+1$. We now consider the soft limit ${\cal Z}_n \to c_- {\cal Z}_{n-1}+c_{+} {{\cal Z}_1}$. For $j=n-1$, this implies
$$[1 \, i \, i+1 \, n-1, n ] \to 0$$
so that we simply get ${\cal P}^{(0)}_{n,NMHV} \to {\cal P}^{(0)}_{n-1,NMHV}$, as expected. As an instructive example, let us consider $n=6$. At tree level, the answer is given by 
\begin{equation}
{\cal P}^{(0)}_{6,NMHV} = [12345]+[13456]+[12356]
\end{equation}
As we take the soft limit, $[13456],[12356] \to 0$. We then get
\begin{equation}
{\cal P}^{(0)}_{6,NMHV} \to  [12345] = {\cal P}^{(0)}_{5,NMHV} = {\cal P}^{(0)}_{5,\overline{MHV}}.
\end{equation}
For $n=5$, the $NMHV$ amplitude agrees with the $\overline{MHV}$ amplitude. 

\subsubsection{One loop}

Let us start with $n=6$. The one-loop result can be written as \cite{Dixon:2011nj}
\begin{eqnarray}
{\cal P}^{(1)}_{6,NMHV} &=&\frac{1}{2}\left([23456]+[12356] \right) V^{(1)}(u_1,u_2,u_3) +\frac{1}{2} \left([13456]+[12346] \right) V^{(1)}(u_2,u_3,u_1) \nonumber \\
&&+\frac{1}{2} \left([12456]+[12345] \right) V^{(1)}(u_3,u_1,u_2).
\end{eqnarray}
Starting from two loops, there is also a parity-odd contribution. The explicit function $V(u_1,u_2,u_3)$ at one loop is given by
\begin{equation}
V^{(1)}(u_1,u_2,u_3)= \frac{1}{2} \left( -\log u_1 \log u_3+ \log(u_1 u_3)\log u_2 -2 \zeta_2 +\sum_{i=1}^3 Li_2(1-u_i) \right)
\end{equation}
The soft limit considered at tree level corresponds to $x_6 \to x_1$. At the level of the cross-ratios, this implies $(u_1,u_2,u_3) \to (1,0,0)$. As explained in the previous section, this limit needs to be taken carefully. We use the parametrisation (\ref{hexagonparam}) and take $\tau,\sigma \to \infty$ with $\tau - \sigma$ fixed. Taking the soft limit, we find
\begin{eqnarray}
{\cal P}^{(1)}_{6,NMHV} \to \frac{1}{2}[12345]\left({\kappa_+} V^{(1)}(u_1,u_2,u_3)+{ \kappa_-} V^{(1)}(u_2,u_3,u_1)+V^{(1)} (u_3,u_1,u_2) \right) \to 0.
\end{eqnarray}
Interestingly, it is important that ${\kappa_+ + \kappa_-=1}$ for the limit to exist. This is, of course, the expected result. Indeed, ${\cal P}^{(1)}_{5,NMHV} = {\cal P}^{(1)}_{5,\overline{MHV}}=0$, since the ratio is defined by dividing by the MHV superamplitude, and the coupling dependence of the MHV and anti-MHV amplitudes is the same. 

The results for ${\cal P}^{(1)}_{n,NMHV}$ for general $n$ can be found in \cite{Elvang:2009ya}. They have the structure
\begin{equation}
{\cal P}^{(1)}_{n,NMHV}  = {\cal P}^{(0)}_{n,NMHV}  P^n_{tot}({\bf u}) + [134 n-1\,n]P^n({\bf u}) +\sum_{s=5}^{n-2}\sum_{t=s+2}^n [1,s-1,s,t-1,t] P^n_{s,t}({\bf u}) + \text{cyclic},
\end{equation}
where $P_{tot}({\bf u}),P({\bf u}),P_{s,t}({\bf u})$ are functions of the overcomplete set of cross-ratios $u_{ijkl}$, eq. (\ref{cr}), denoted collectively by ${\bf u}$. The results are written in a cyclically symmetric way, but they involve an overcomplete basis of $R$-invariants as well. We can use the linear relations among them to write the result in terms of a linearly independent set; for instance, \cite{Elvang:2010xn}
\begin{equation}
{\cal P}^{(1)}_{n,NMHV} = \sum_{1 < j <k < \ell< m \leq n} [1, j, k, \ell,m] V^{n}_{j k \ell m}({\bf u}).
\end{equation}
We now take the soft limit. At the level of the $R$-invariants, this amounts to
\begin{equation}
[1,j,k,n-1,n] \to 0,~~~ [1,j,k,\ell,n] \to {\kappa_-} [1,j,k,\ell,n-1] ~~\text{for $\ell<n-1$}
\end{equation}
while at the level of coordinates, this amounts to $x_n \to x_1$, which implies $x^2_{n-1,1} \to 0$, $x_{n,2}^2 \to 0$ and $x^2_{\ell n} \to x^2_{1,\ell}$. In all cases explicitly checked we find
\begin{eqnarray}
V^n_{jk\ell n}({\bf u}) \to 0,~~~\text{for $\ell<n-1$},\\
V^n_{jk\ell m}({\bf u}) \to V^{n-1}_{jk\ell m}({\bf u}'),~~~\text{for $m<n$},
\end{eqnarray}
where ${\bf u}'$ is the reduced set of cross-ratios for the scattering of $n-1$ particles. In particular, this implies
\begin{equation}
{\cal P}^{(1)}_{n,NMHV} \xrightarrow{\text{soft}}  {\cal P}^{(1)}_{n-1,NMHV}
\end{equation}
We have checked this directly for the full answer of \cite{Elvang:2009ya} for very high $n$. Interestingly, one does not need to use the (highly nonlinear!) relations among the cross-ratios. Only the relations among the $R$-invariants are necessary.

\subsection{Half-collinear behaviour}

For the discussion on the $S-$algebra below, we need to consider the holomorphic half-collinear limit between two adjacent positive-helicity gluons. In terms of the ratio ${\cal P}_n$ this means we need to set $\eta_j=\eta_{j+1}=0$ for the two adjacent particles. With this restriction, at tree-level we obtain
\begin{equation}
{\cal P}^{(0)}_n(\cdots,p_j,p_{j+1},\cdots)\xrightarrow{\text{half-coll}}
{\cal P}^{(0)}_{n-1}(\cdots,p,\cdots).
\label{half-coll-tree}
\end{equation}
To see this, consider for example (\ref{tree-level-P}), and take particles $1,n$ to become half-collinear, so that $\lambda_1 \parallel \lambda_n$. In this limit the four-brackets $\langle 1 \, a \, b \, n \rangle$ are suppressed respect to the other brackets. After projecting the two collinear legs to positive-helicity gluons
\begin{equation}
\left.[1\,a\,b\,c\,n]\right|_{\eta_1=\eta_n=0}
\xrightarrow{\langle n1\rangle\to0}0,
\end{equation}
from which (\ref{half-coll-tree}) follows. At finite coupling the argument is the supersymmetric analogue of section \ref{sec:collinearnp}. In the variables of that section, the limit corresponds to $z_s\to z_2$ with $\bar z_s,\omega_s$ fixed, or equivalently $i\phi\equiv \phi_e$ with $\tau,\phi_e\to\infty$ and $\tau-\phi_e,\sigma$ fixed. By the superamplitude/super Wilson-loop duality \cite{Caron-Huot:2010ryg,Mason:2010yk}, each Grassmann component of ${\cal P}_n$ admits a superloop OPE expansion \cite{Sever:2011da,Dixon:2011nj,Dixon:2014iba}. In the channel where the edges carrying $s$ and $2$ become collinear, the expansion has the schematic form
\begin{equation}
{\cal P}_n(\tau,\sigma,\phi,{\bf v};\chi)
=
{\cal P}_{n-1}({\bf v};\chi)
+
\sum_{\Psi\neq 0}
C_\Psi(\sigma,{\bf v};\chi)\,
e^{-E_\Psi(a)\tau+i p_\Psi(a)\sigma+i m_\Psi\phi},
\label{eq:P-super-OPE}
\end{equation}
where the first term is the vacuum exchange and the sum runs over non-vacuum super-flux-tube states. The Grassmann variables only dress the OPE coefficients; the dependence on $\tau,\sigma,\phi$ is governed by the same flux-tube quantum numbers that appear in the bosonic Wilson-loop OPE. At finite coupling the GKP spectrum obeys the strict gap $E_\Psi(a)-|m_\Psi|>0$ for every non-vacuum exchange \cite{Basso:2010in}, just as in section \ref{sec:collinearnp}. Therefore in the half-collinear limit each term in the sum is bounded by
\begin{equation}
e^{-E_\Psi\tau+i m_\Psi\phi}
\sim
e^{-(E_\Psi-|m_\Psi|)\tau}
\times
\text{finite},
\end{equation}
and vanishes, while the vacuum term survives. This establishes, at finite coupling, the half-collinear behaviour
\begin{equation}
\boxed{\displaystyle
{\cal P}_n(\cdots,s,2,\cdots)\xrightarrow{\text{half-coll}}
{\cal P}_{n-1}(\cdots,p,\cdots).
}
\label{eq:P-half-coll}
\end{equation}
Hence the half-collinear splitting of ${\cal P}_n$ for two consecutive positive-helicity gluons is uncorrected from its tree value, and together with the corresponding result for $R_n$ it follows that the leading half-collinear pole of $\mathcal A_n^{\rm hard}$ is exactly the tree-level one.

\section{Soft/hard factorization}
\label{sec:IRfactorization}

The full color-ordered superamplitude can be written in a soft/hard 
factorized form
\begin{equation}
\mathcal{A}_n \;=\; \mathcal{A}^{\rm soft}_n(\epsilon)\;\mathcal{A}^{\rm hard}_n \, ,
\label{eq:soft-hard-factorization}
\end{equation}
where $\mathcal{A}^{\rm soft}_n$ contains all IR divergences, while $\mathcal{A}^{\rm hard}_n$ is IR finite and
hence has a smooth $\epsilon\to0$ limit. There are many ways to implement this soft factorization.\footnote{Other examples of choices appear in \cite{Gonzalez:2021dxw,Bhardwaj:2022anh}.} Here we choose
\begin{equation}
\boxed{
\;\mathcal{A}^{\rm soft}_n(\epsilon)\equiv  e^{G_n(\epsilon)}\,,
\qquad
\mathcal{A}^{\rm hard}_n \equiv \mathcal{A}^{\rm tree}_{n,\rm MHV}e^{R_n}\,{\cal P}_n \; }\, ,
 \label{eq:def-soft-hard}
\end{equation}
with 
\be G_n\equiv 
\log M_n-R_n .\ee
$G_n(\epsilon)$ governs the soft sector and is known fairly explicitly:
 \bea 
 \label{eq:Gn-soft-factor}
 G_n=
  -\frac{1}{4} \sum_{\ell=1}^\infty a^\ell \left( \frac{f^{(\ell)}}{(\ell \epsilon)^2} + \frac{g^{(\ell)}}{(\ell \epsilon)}   \right) \sum_{i=1}^n \left( \frac{-s_{i,i+1}}{\mu_{IR}^2}\right)^{-\ell \epsilon} +\frac{1}{2} f(a) F^{(1-loop)}_n +h_n(a) +\cal{O}(\epsilon).
\eea
We also recall
\begin{equation}
\mathcal{A}^{\rm tree}_{n,\rm MHV}
=
i\,\frac{\delta^{(4)}\!\left(\sum_{i=1}^n p_i\right)\,\delta^{(8)}\!\left(\sum_{i=1}^n \lambda_i \eta_i\right)}
{\langle 12\rangle\langle 23\rangle\cdots\langle n1\rangle}\, . 
\label{eq:MHV-tree-times-Mn}
\end{equation}
The advantage of this specific division is that, under the assumptions and soft-limit results above,  $\mathcal{A}^{\rm hard}_n$ obeys the uncorrected tree-level leading soft theorem\footnote{Despite the name, there is no restriction on how soft the scattering states in $\mathcal{A}^{\rm hard}_n$ may be taken.} 
\be \label{sft}
\lim_{s \to 0} \mathcal{A}^{\rm hard}_{n+1}(1,\cdots n,s)= S^{(0)}\mathcal{A}^{\rm hard}_{n}(1,\cdots n)+\cdots,~~~~~~S^{(0)}=\frac{\langle n,1\rangle}
{\langle n,s\rangle\langle s,1\rangle},\ee
while still encoding most of the information about higher loop corrections. Moreover,  for two adjacent positive-helicity gluons, its splitting in the collinear limit is given exactly by the tree-level result 
\begin{equation}\label{splt}
 \lim_{p_j \parallel p_{j+1}}\mathcal{A}^{\rm hard}_{n+1}(\cdots, p_j,p_{j+1},\cdots) = \frac{1}{\sqrt{z(1-z)}\langle p_j ,p_{j+1}\rangle }\mathcal{A}^{\rm hard}_n(\cdots,p,\cdots) +\cdots~~~~~~p=p_j +p_{j+1}.
\end{equation} 
Hence, despite the fact that $\mathcal{A}^{\rm hard}$ captures much of the quantum corrections to scattering, its leading positive-helicity soft-collinear behavior is tree-level exact. 

A further related property is that $\mathcal{A}^{\rm hard}$ is covariant under dual conformal transformations, without any anomaly, since it is given by the product of the tree-level amplitude, which is covariant, and a dual conformally invariant piece. More specifically,
\begin{equation}
K^\mu  \mathcal{A}^{\rm hard}_n=-2\sum_i x_i^\mu \mathcal{A}^{\rm hard}_n,~~~D \mathcal{A}^{\rm hard}_n=-n \mathcal{A}^{\rm hard}_n
\end{equation}
where $K^\mu,D$ are the full dual superconformal generators, including their action on the spinor and Grassmann variables, see Appendix B of \cite{Drummond:2008vq}.

The IR divergent part of one-loop corrections to the subleading soft theorem were studied by Bern, Davies and Nohle (BDN) in \cite{Bern:2014oka}.  It is instructive to see how their results are consistent with ours.   In the notation of BDN, write $G_m=aG_m^{[1]}+{\cal O}(a^2)$ and define
\begin{equation}
\sigma_m(\epsilon)\equiv\frac12\sum_{i=1}^{m}
\left(\frac{-s_{i,i+1}}{\mu_{\rm IR}^2}\right)^{-\epsilon},
\qquad (G_m^{[1]})_{\rm div}=-\frac{\sigma_m}{\epsilon^2}.
\end{equation}
When leg $n$ is soft,
\begin{equation}
\sigma_n'=\frac12\left[
\left(\frac{-s_{n-1,n}}{\mu_{\rm IR}^2}\right)^{-\epsilon}
+\left(\frac{-s_{n,1}}{\mu_{\rm IR}^2}\right)^{-\epsilon}
-\left(\frac{-s_{n-1,1}}{\mu_{\rm IR}^2}\right)^{-\epsilon}
\right],\qquad \sigma_n=\sigma_{n-1}+\sigma_n'.
\end{equation}
After converting conventions, $g_{\rm YM}^2N c_\Gamma=a/2+{\cal O}(\epsilon^2)$. For stripped amplitudes evaluated with the common $n$-point kinematic prescription of BDN, conjugating the hard subleading operator with the soft factor gives
\begin{equation}
S^{(1)}_{\rm full}
=e^{G_n}S^{(1)}_{\rm hard}e^{-G_{n-1}}
=S^{(1)}_{\rm hard}
-\frac{a}{\epsilon^2}\left[\sigma_n' S^{(1)}_{\rm tree}-\left(S^{(1)}_{\rm tree}\sigma_{n-1}\right)\right]
+{\cal O}(a\epsilon^0,a^2).
\end{equation}
With this convention conversion, 
the square bracket is precisely the divergent one-loop subleading correction found in \cite{Bern:2014oka}. Thus the BDN term is reproduced by the soft variation of $\mathcal A^{\rm soft}$, while any remaining correction to the subleading soft theorem of $\mathcal A^{\rm hard}$ is IR finite by construction. Hence the one loop anomaly found by BDN is not an obstruction to $S$-covariance of IR finite quantities.  Related observations were made in \cite{Cachazo:2014dia,He:2017fsb}.  Finite corrections to the subleading soft theorem were computed at one-loop in \cite{Brandhuber:2015vhm}. Partial results to higher orders can be found in the Appendix. 

\section{$S$-algebra}
\label{sec:undeformedS}

In this section, we argue that, even though the SYM soft theorems themselves may (or may not) have loop corrections beyond leading order, the commutator algebra of the tower of soft theorems obeyed by $\mathcal{A}^{\rm hard}$ is the undeformed tree-level $S$-algebra. We expect similar conclusions to pertain to the ${\cal N}=4$ extension of $S$, but we shall not consider it here.

 The soft theorems relate two distinct objects. First, there are soft gluon insertions in an ($n+1$)--point amplitude, denoted by $\mathsf S$. Their short-distance OPE algebra is read off of the half-collinear splitting function. Second, there are the differential operators that appear on the right-hand side of a soft theorem, denoted by $\mathsf T$; these act on the $n$--point amplitude with the soft insertion removed. We will 
 see that the collection of all amplitudes of fixed $n$ forms a representation (or module) of $\mathsf T$, while $\mathsf S$ is an intertwiner connecting different representations. 
 \subsection{Soft Algebras}
 The tower of $\mathsf S$  generators and their commutators can be described either in plane wave basis of energy eigenstates, or in a Lorentz/conformal basis of boost eigenstates. At loop level, the energy basis becomes hard to define due to $\log \omega$ corrections. The analysis appears simpler in the conformal basis, where the leading holomorphic singularity in the gluon OPE is just the Mellin transform of this collinear pole in \eqref{splt} \cite{Guevara:2021abz,Strominger:2021mtt}. We briefly recap this construction here. The boost-weight-$\Delta$, positive-helicity, outgoing conformal primary gluon operator is the Mellin transform of the annihilation operator
\begin{equation}
  \mathcal O^a_{\Delta,+}(z,\bar z)
  =
  \int_0^\infty d\omega\,\omega^{\Delta-1}\,
  a^a_+(\omega,z,\bar z),
\label{eq:celestial-mellin-O}
\end{equation}
where $a$ is a color index and the map $p_i \to (\omega_i,z_i,\bar z_i)$ is given in (\ref{ptow}).  IR finite celestial amplitudes in the conformal basis, denoted $\mathsf{A}_n^{\rm hard}$, are obtained by applying this Mellin transform to every external energy of $\mathcal{A}^{\rm hard}$:
\be \mathsf{A}_n^{\rm hard}(\Delta_1,z_1,\bz_1; \ldots \Delta_n,z_n,\bz_n) =\prod_{k=1}^n\int_0^\infty d\omega_k\,\omega_k^{\Delta_k-1}\,\mathcal{A}^{\rm hard}(p_1;\ldots p_n).\ee
OPEs of the $\mathcal O^a_{\Delta,+}$ are then obtained from the collinear expansion of the hard celestial amplitudes. Since the leading pole in the splitting function is unchanged, the leading OPE is given by the tree-level result 
\begin{equation}
\begin{aligned}
  \mathcal O^a_{\Delta_1,+}(z_1,\bar z_1)\,
  \mathcal O^b_{\Delta_2,+}(z_2,\bar z_2)
  \sim{}&
  -\frac{i f^{ab}{}_{c}}{z_{12}}
  \sum_{r=0}^{\infty}
  B(\Delta_1-1+r,\Delta_2-1)\\
  &\times\frac{\bar z_{12}^{r}}{r!}\,
  \bar\partial_2^r
  \mathcal O^c_{\Delta_1+\Delta_2-1,+}(z_2,\bar z_2)
  +O(z_{12}^{0}).
\end{aligned}
\label{eq:hard-celestial-ope}
\end{equation}
All displayed antiholomorphic descendants remain at order $1/z_{12}$ and are needed for the mode algebra.
 Loop effects can still change less singular terms in the OPE, or equivalently the detailed action of subleading soft operators on hard data, but they do not change the pole which fixes the insertion-mode algebra.\footnote{Indeed, as argued in section \ref{sec:collinearnp}, loop effects at finite coupling give a contribution suppressed by a strictly positive power of $z_{12}$ in the OPE limit.}

The beta function in (\ref{eq:hard-celestial-ope}) has conformally soft poles at every integer value $\Delta\le 1$. It is conventional to label this tower by
\begin{equation}
p=\frac{3-\Delta}{2}=1,\frac{3}{2},2,\frac{5}{2},\ldots .
\label{eq:soft-delta-tower}
\end{equation}
The leading soft current discussed above has $p=1$ and $\Delta=1$. We define the gamma-normalized generators $\mathsf S^{p,a}_{\bar m,m}$ as modes of $\mathcal O^a_{\Delta,+}$\footnote{The contour integrals here are at fixed $\bar z$, which is possible for complexified momenta or real momenta in Klein space.}\begin{equation}
 \begin{aligned}
  \mathsf S^{p,a}_{\bar m,m}
  &=
  \Gamma(p-\bar m)\Gamma(p+\bar m)
  \oint\frac{dz}{2\pi i}\,z^{m+1-p}
  \oint\frac{d\bar z}{2\pi i}\,\bar z^{\bar m-p}\,
\oint \limits_{\Delta=3-2p} \frac{d\Delta}{2\pi i}   \mathcal O^a_{\Delta,+}(z,\bar z),
 \end{aligned}
\label{eq:S-from-Odelta}
\end{equation}
where the contours are around the poles and the residues vanish outside the wedge $|\bar m|<p$. The leading pole in the OPE algebra 
then gives the undeformed soft insertion algebra 
\begin{equation}
  [\mathsf S^{p,a}_{\bar m,m},\mathsf S^{q,b}_{\bar n,n}]
  =
  -i f^{ab}{}_{c}\,
  \mathsf S^{p+q-1,c}_{\bar m+\bar n,m+n}.
\label{eq:undeformed-S-algebra}
\end{equation}
For $p=q=1$, the only anti-holomorphic weight is $\bar m=\bar n=0$, and (\ref{eq:undeformed-S-algebra}) reduces to the leading current algebra that follows immediately from the splitting function. The point of (\ref{eq:undeformed-S-algebra}) is stronger: the same tree-level insertion algebra holds for the whole tower $\Delta=1,0,-1,\ldots$. This is the $S$-algebra in the sense of \cite{Guevara:2021abz,Strominger:2021mtt,Kmec:2025ftx}; it is not just the leading $p=1$ current. The important point for us is that the structure constants in this insertion algebra are not renormalized.  

At tree level, the leading soft current together with special conformal symmetry generates the entire conformally soft tower \cite{Larkoski:2014hta,Sheta:2025oep} and implies the $S$-algebra. In the present context, the usual special conformal symmetry is anomalous at loop level, and this argument cannot be directly applied. Nevertheless, we see. here that the same conclusions can be reached using the nonrenormalization of the splitting function defined by $\mathcal{A}^{\rm hard}$.

\subsection{Soft theorems}
Now we turn to soft theorems. These relate amplitudes with two different multiplicities: an $(n+1)$--point amplitude with a conformally soft gluon $\mathsf S$ equals the action of $\mathsf T$ on an $n$--point amplitude. Schematically,
\begin{equation}
  \left\langle \mathsf S^{p,a}_{\bar m,m}\,\prod_i \mathcal O_i\right\rangle_{\rm hard}
  =\mathsf T^{p,a}_{\bar m,m}\,
  \left\langle \prod_i \mathcal O_i\right\rangle_{\rm hard}.
\label{eq:soft-theorem-action}
\end{equation}
In a general setting, especially for $p>1$, $\mathsf T$ can receive loop-dependent corrections: \begin{equation}
  \mathsf T^{p,a}_{\bar m,m}
  =
  \mathsf T^{p,a,{\rm tree}}_{\bar m,m}
  +\Delta \mathsf T^{p,a}_{\bar m,m}.
\label{eq:S-deformed-action}
\end{equation}
The tree-level term is a differential operator in $\bar z$ of degree $2p-2$ and lowers the conformal weight by $2p-2$. An explicit expression may be obtained by commuting the leading operator with special conformal transformations \cite{Larkoski:2014hta,Sheta:2025oep}. We define
\begin{equation}
  \bar{\mathcal D}_i(z,\bar z)
  \equiv 
  \Bigl((\bar z-\bar z_i)\bar\partial_i-2\bar h_i+1\Bigr)
  e^{-\partial_{\Delta_i}} .
\label{eq:Dbar-def}
\end{equation}
and
\begin{equation}
  \mathsf T^{p,a,{\rm tree}}(z,\bar z)
  =
  {1\over\Gamma(2p-1)}
  \sum_i T_i^a\,{1\over z-z_i}\,
  \bar{\mathcal D}_i(z,\bar z)^{2p-2}.\footnote{The power denotes ordinary operator composition, with the rightmost factor acting first.  On leg $i$, let $E$ denote the unit downward shift of $\Delta$ and write $N=2p-2$ and $X=(\bar z-\bar z_i)\bar\partial_i-2\bar h_i+1$.  Then $EX=(X+1)E$, because $E$ shifts $\bar h_i=(\Delta_i-J_i)/2$ by $-\frac12$.  Together with $\bar\partial_i(\bar z-\bar z_i)=-1$, this gives $\bar{\mathcal D}_i^{N}=(X)_N E^N$, where $(X)_N=X(X+1)\cdots(X+N-1)$.  Normal ordering produces a finite Pochhammer sum.}
\label{eq:Tp-Dbar-power}
\end{equation}
Then one finds that $T^{p,a,{\rm tree}}_{\bar m,m}$ are the mode coefficients of $\mathsf T^{p,a,{\rm tree}}(z,\bar z)$. Explicitly,
\be \mathsf T^{p,a,{\rm tree}}_{\bar m,m}
  =
  \Gamma(p-\bar m)\Gamma(p+\bar m)
  \oint\frac{dz}{2\pi i}\,z^{m+1-p}
  \oint\frac{d\bar z}{2\pi i}\,\bar z^{\bar m-p}\,
  \mathsf T^{p,a,{\rm tree}}(z,\bar z).\ee
In the previous subsection, we argued that the soft insertions on the left-hand side of \eqref{eq:soft-theorem-action} obey an uncorrected $S$-algebra. This implies that the full $\mathsf T$ action on the right-hand side obeys the same uncorrected algebra.  The explicit form of the tree-level generators \eqref{eq:Tp-Dbar-power} then implies integrability conditions on any corrections $\Delta \mathsf T$. Explicit checks of this would be of interest.

\section{Discussion}

We have shown that, given stated assumptions,  color-ordered amplitudes in planar ${\cal N}=4$ SYM admit a factorized form $\mathcal{A}_n \;=\; \mathcal{A}^{\rm soft}_n(\epsilon)\;\mathcal{A}^{\rm hard}_n$, where $\mathcal{A}^{\rm hard}_n$ is IR-finite and obeys the uncorrected tree-level leading soft theorem.  This result is  the starting point for a discussion of soft algebras and theorems for planar ${\cal N}=4$ SYM, and there are many interesting open questions. 

\begin{itemize}
\item Can we understand $\mathcal{A}^{\rm soft}_n(\epsilon)$ as IR dressings or a soft-sector CFT$_2$ along the lines of \cite{Arkani-Hamed:2020gyp,Magnea:2021fvy,Gonzalez:2021dxw,Choi:2024ygx,Choi:2026cyh,Choi:2026dhz}? In principle, one may hope to find a 2D theory which fully reproduces it, as was done for the much simpler cases of QED and gravity. A related goal would be to understand the IR-finite $\mathcal{A}^{\rm hard}_n$ as the amplitudes for suitably dressed states. 
\item The leading soft theorems agree with those of the tree-level amplitudes. Subleading theorems, on the other hand, appear to be deformed at loop level, and in the appendix we have computed the leading logarithmic contribution. This deformation should be closely connected to the holomorphic anomaly, which causes the usual special conformal transformations to be anomalous at loop order \cite{Beisert:2010gn}. Indeed, at tree level, special conformal transformations together with the leading soft theorem generate the tower of soft theorems \cite{Larkoski:2014hta,Sheta:2025oep}. It would be very interesting to better understand both the anomaly of special conformal transformations in planar ${\cal N}=4$ SYM and its connection to subleading soft theorems. 
\item Subleading soft theorems in momentum space contain logarithms of the soft momentum. It would be interesting to study the effect of these logarithms in the celestial basis, where logs translate into higher poles. The celestial basis has been instrumental in finding the $S$-algebra, and we expect it to play an important role in understanding potential deformations. The recent discussion \cite{Sheta:2025oep} gives another useful perspective on how 4D conformal symmetry organizes the tower.
\item The hard amplitude gives evidence that the $S$-algebra itself is not deformed, even though the subleading soft theorem can be. This distinction is important in comparing with examples where soft algebras are genuinely deformed by additional interactions/backgrounds \cite{Melton:2022fsf,Fernandez:2023kdu,Costello:2022aol,Taylor:2023ajd,Costello:2022wso}. 

\item In some examples, e.g., \cite{Sheta:2025oep}, 4D commutators, 2D commutators, and double-soft limits are all equal or proportional. It would be very interesting to derive the undeformed closure directly from double-soft limits of $\mathcal{A}^{\rm hard}_n$.
\item It would be very interesting to develop the OPE for Wilson loops into a tool to study soft corrections to scattering amplitudes in planar ${\cal N}=4$ SYM. In the present paper we have barely scratched the surface of this. 
\end{itemize}

\section*{Acknowledgements}

We are grateful to Nima Arkani-Hamed, Benjamin Basso, and Lance Dixon for useful conversations. OpenAI internal models and GPT-5.5 were used in all stages of this work. A.S.'s work was partially supported by the Simons Collaboration for Celestial Holography, the Moore Foundation via the Black Hole Initiative, and DOE grant DE-SC/0007870. L.F.A.'s work is partially supported by the STFC grant ST/T000864/1. For the purpose of open access, the authors have applied a CC BY public copyright licence to any Author Accepted Manuscript (AAM) version arising from this submission.

\appendix

\section{Corrections to the soft limit}
In this appendix, we give further details of the construction in \cite{Alday:2010ku,Basso:2013vsa,Basso:2013aha,Basso:2014koa}, and use it to compute the leading nontrivial correction to the remainder function in the soft limit. In \cite{Brandhuber:2015vhm}, the constraints arising from dual conformal symmetry were analysed at one loop, and a simple expression was found for the logarithmic contributions. Here we would like to analyse some of the constraints arising from the WL OPE, where dual conformal invariance is built in. We will focus on the simplest nontrivial contribution at each order in perturbation theory, namely the term proportional to $\delta \, a^L \log^L\delta$. As we will see, this term has a universal form that can be predicted to all loops in perturbation theory. At one loop, we will reproduce the results found in \cite{Brandhuber:2015vhm}. 

\subsection{Wilson loop multi-channel OPE}

The general idea of the WL OPE was explained in the body of the paper, where the leading soft limit was considered. To compute corrections to this leading behaviour, we will use a generalization of those ideas developed in \cite{Basso:2013vsa,Basso:2013aha,Basso:2014koa}. 

\begin{figure}{}
\centering
 \includegraphics[width=0.2\textwidth]{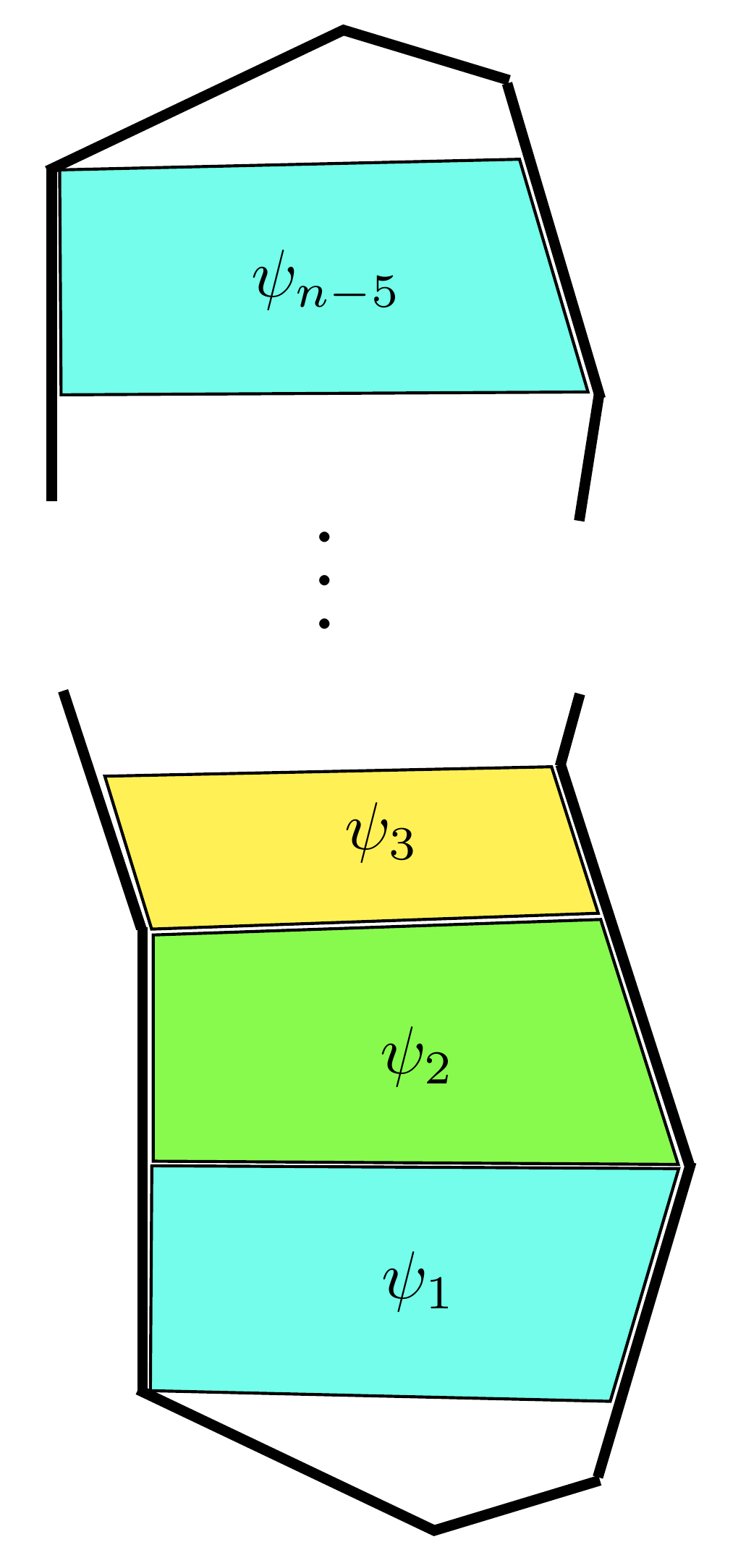}
\caption{Tessellation of ${\cal W}_n$ into $n-5$ reference squares}
\label{figa1}
\end{figure}
We start with a polygonal null Wilson loop with $n$ sides ${\cal W}_n$. We then tessellate it into $n-5$ reference squares, see figure \ref{figa1}, whose expectation values we denote by ${\cal W}^{(1)}_4,\cdots, {\cal W}^{(n-5)}_4$. Two consecutive squares form a pentagon. Together with the two pentagons at the bottom and at the top, we have $n-4$ pentagons, whose expectation values we denote by ${\cal W}^{(1)}_5,\cdots,{\cal W}^{(n-4)}_5$. We then consider the ratio
\begin{equation}
\hat {\cal R}_n = {\cal W}_n \frac{{\cal W}^{(1)}_4 \times \cdots \times {\cal W}^{(n-5)}_4}{{\cal W}^{(1)}_5 \times \cdots \times  {\cal W}^{(n-4)}_5 }.
\end{equation}
This ratio is finite and dual conformal invariant (or simply conformal invariant in the Wilson-loop language). It is closely related to the ratio ${\cal R}_n$ introduced in the body of the paper, but slightly better suited for our current purposes. As in the single-channel OPE, $\hat {\cal R}_n$ admits a decomposition in terms of eigenstates of the reference squares, of the form
\begin{equation}
\hat {\cal R}_n  = \sum_{\psi_1,\cdots,\psi_{n-5}} {\cal P}(0| \psi_1) {\cal P}( \psi_1| \psi_2) \cdots  {\cal P}( \psi_{n-6}| \psi_{n-5}) {\cal P}( \psi_{n-5}| 0) e^{\sum \limits_{j=1}^{n-5} -E_j \tau_j +i p_j \sigma_j+i m_j \phi_j}.
\end{equation}
We start with the vacuum at the bottom and evolve it all the way up to the vacuum at the top. In between, we decompose the flux-tube state in the $i$th middle square into a basis of eigenstates $\psi_i$, with energy $E_i$, momentum $p_i$, and $SO(2)$ charge $m_i$, each weighted by the appropriate exponential factor. Here $\tau_i,\sigma_i,\phi_i$ are cross-ratios associated with the symmetries of the $i$th square. They are three independent cross-ratios that parametrize the hexagon formed by the two pentagons sharing the $i$th square. In total, we have $3(n-5)$ independent cross-ratios, as expected. The pentagon transitions ${\cal P}( \psi_i| \psi_{i+1})$ correspond to pentagon Wilson loops with two states $\psi_i, \psi_{i+1}$ attached to non-adjacent sides. 

The states are generally $N$-particle states. For our purposes\footnote{We are interested in computing the leading logarithmic behaviour of the leading soft corrections at each order in perturbation theory.}, however, it will be enough to focus on single-particle states. These are parametrised by a rapidity $u$ together with a discrete label that indicates the type of excitation; see \cite{Basso:2010in}. For the case of MHV amplitudes, which involve bosonic loops, the states have to be singlets under $SU(4)_R$. The excitations relevant for us are gauge-field excitations $F_{\pm}$, with $SO(2)$ charge $\pm 1$, and their bound states $D^{k-1}_\pm F_\pm$, with $SO(2)$ charge $\pm k$. Using this more refined notation, we have, for instance,
\begin{align*}
 &\hat {\cal R}_6 \! = \! \sum_{k} \! \int \! \frac{du}{2\pi}\mu_k(u) P_{k}(0|u) P_{k}(-u|0) e^{-E_k(u)\tau_1 +i p_k(u)\sigma_1 + i k \phi_1} \cr
 &\hat {\cal R}_7\! = \! \sum_{k,k'} \! \int \! \frac{du}{2\pi} \frac{dv}{2\pi}\mu_k(u) \mu_{k'}(v) P_{k}(0|u) P_{k k'}(-u|v) P_{k'}(-v|0) e^{-E_k(u)\tau_1 +i p_k(u)\sigma_1 + i k \phi_1} e^{-E_{k'}(v)\tau_2 +i p_{k'}(v)\sigma_2 + i {k'} \phi_2} 
\end{align*}
and so on, where in this notation $k,k'$ can be positive or negative. Here $\mu_k(u)$ is the measure and $P_{kk'}(-u|v)$, etc., give the pentagon transitions. Their computation as functions of the coupling constant has been the subject of the developments in \cite{Basso:2013vsa,Basso:2013aha,Basso:2014koa}. 

Let us denote $\{ \tau_1,\sigma_1,\phi_1\} \equiv \{ \tau,\sigma,\phi\}$, since the first set of cross-ratios plays a prominent role in what follows. We work with the normalisation where the creation amplitude for a single particle is one, $P_{k}(0|u)=P_{k}(-u|0)=1$. We furthermore introduce the following notation:
\begin{equation}
\hat {\cal R}_{n}(\sigma,\tau,\phi) = \sum_k \int \frac{du}{2\pi} \mu_k(u) e^{- E_k(u)\tau +i p_k(u) \sigma + i k \phi}\hat {\cal R}^{(k,u)}_{n-1},
\end{equation}
where $\hat {\cal R}^{(k,u)}_{n-1}$ corresponds to a Wilson loop with $n-1$ edges, with an extra insertion of the state labelled by $(k,u)$ at its bottom (instead of the vacuum). The vacuum corresponds to $k=0$, in which case $\hat {\cal R}^{(0)}_{n-1}=\hat {\cal R}_{n-1}$. Both $\hat {\cal R}_{n}(\sigma,\tau,\phi)$ and $\hat {\cal R}_{n-1}$ depend on the extra $3(n-6)$ cross-ratios, but we are suppressing this dependence. 

We take the soft limit by sending $\tau,\sigma \to \infty$ with $\tau - \sigma$ fixed. We do this in two steps. First, we consider the collinear limit, which corresponds to $\tau \to \infty$. The expansion around the collinear limit is controlled by the energy of the first state $\psi_1$. The state with the lowest energy is the vacuum. This leads to the leading collinear (and soft) limit $\hat {\cal R}_n \to \hat {\cal R}_{n-1}$. The leading correction arises from the exchange of gluons $F_{\pm}$, with $m_{\pm}=\pm 1$, whose energy in perturbation theory is given by $E(u) = 1 +\epsilon(u)$. We obtain
\begin{equation}
\hat {\cal R}_n(\sigma,\tau,\phi) = \hat {\cal R}_{n-1} +\sum_{\pm} e^{-\tau} e^{\pm i \phi}  \int_{-\infty}^\infty \frac{du}{2\pi} \, \mu(u) e^{-\epsilon(u) \tau +i p(u) \sigma} \hat {\cal R}_{n-1}^{(\pm,u)} + \cdots
\label{Rsl}
\end{equation}
In perturbation theory $p(u)=2u+\cdots$ while
\begin{equation}
\epsilon(u)=  \left(H\left(\frac{1}{2}+i u\right)+H\left(\frac{1}{2}-i u\right)\right) a+ \cdots,
\end{equation}
with $H(x)$ the harmonic number. The gluonic measure is given by
\begin{equation}
\mu(u) = -\frac{\pi}{2(u^2+\frac{1}{4}) \cosh(\pi u)} a + \cdots
\end{equation}
Now we consider the soft limit $\sigma \to \infty$. Note that the gluonic measure has poles along the imaginary axis in the $u$-plane. We can compute the leading contribution in the soft limit by closing the contour and picking the contribution from the pole at $u=i/2$. At higher orders in perturbation theory, one should also take into account $\epsilon(u),p(u)$ and $P_{\pm,k'}(-u|v)$, which also contain poles at $u=i/2$. The order of the pole increases, and the general structure of the relevant integral at order $a^{\ell}$ is
\begin{equation}
2\pi i \text{Res}_{u= \frac{i}{2}} \, \frac{\mu(u)}{2\pi} e^{-\epsilon(u) \tau +i p(u) \sigma} P_{\pm,k'}(-u|v) = e^{-\sigma} f_{k'}^{(\ell)}(\sigma,\tau;v),
\end{equation}
with $f_{k'}^{(\ell)}(\tau,\sigma;v)$ a polynomial of degree $\ell$ in $\sigma,\tau$. In general, the dependence on $v$ is very complicated. Here we will consider the simplest contribution: the leading terms of homogeneous order $\ell$ in $\sigma$ and $\tau$. Using the different pieces given in \cite{Basso:2013vsa,Basso:2013aha,Basso:2014koa} to very high order, we were able to guess this leading contribution to arbitrary order in perturbation theory. It is given by 
\begin{equation}
f^{(\ell)}(\sigma,\tau;v) = P_{\rm LL}^{(\ell)}(\sigma,\tau)+ \cdots,~~~ P_{\rm LL}^{(\ell)}(\sigma,\tau)=\frac{(-1)^{\ell} \sigma  \tau ^{\ell-1} \, _2F_1\left(1-\ell,2-\ell;2;\frac{\sigma }{\tau }\right)}{\Gamma (\ell)} 
\end{equation}
Here, $\cdots$ involves contributions suppressed by powers of $\sigma$ and $\tau$ in the large-$\sigma,\tau$ limit. For the first few cases, we obtain
\begin{equation}
P_{\rm LL}^{(1)}(\sigma,\tau) = -\sigma,~~~P_{\rm LL}^{(2)}(\sigma,\tau) = \sigma \tau,~~~P_{\rm LL}^{(3)}(\sigma,\tau) =-\frac{1}{2} \sigma \tau (\sigma+\tau),\cdots
\end{equation}
These are proportional to the Narayana polynomials. An important property is that they are symmetric for $\ell \geq 2$. We then get the following multiplicative structure for the leading terms in the subleading soft expansion, to all orders in perturbation theory:
\begin{equation}
\hat {\cal R}_n(\sigma,\tau,\phi) = \hat {\cal R}_{n-1} + e^{-\tau -\sigma} 2\cos \phi P_{\rm LL}(\sigma,\tau) \hat {\cal R}_{n-1}+ \cdots,
\end{equation}
with $P_{\rm LL}(\sigma,\tau)= \sum_{\ell=1} P_{\rm LL}^{(\ell)}(\sigma,\tau) a^\ell$. At each order in perturbation theory, this contains the leading contribution in the large-$\sigma,\tau$ expansion. Next we translate these results to a more familiar language. 


\subsection{Geometric construction}

\begin{figure}[h]
\centering
\includegraphics[width=0.4\textwidth]{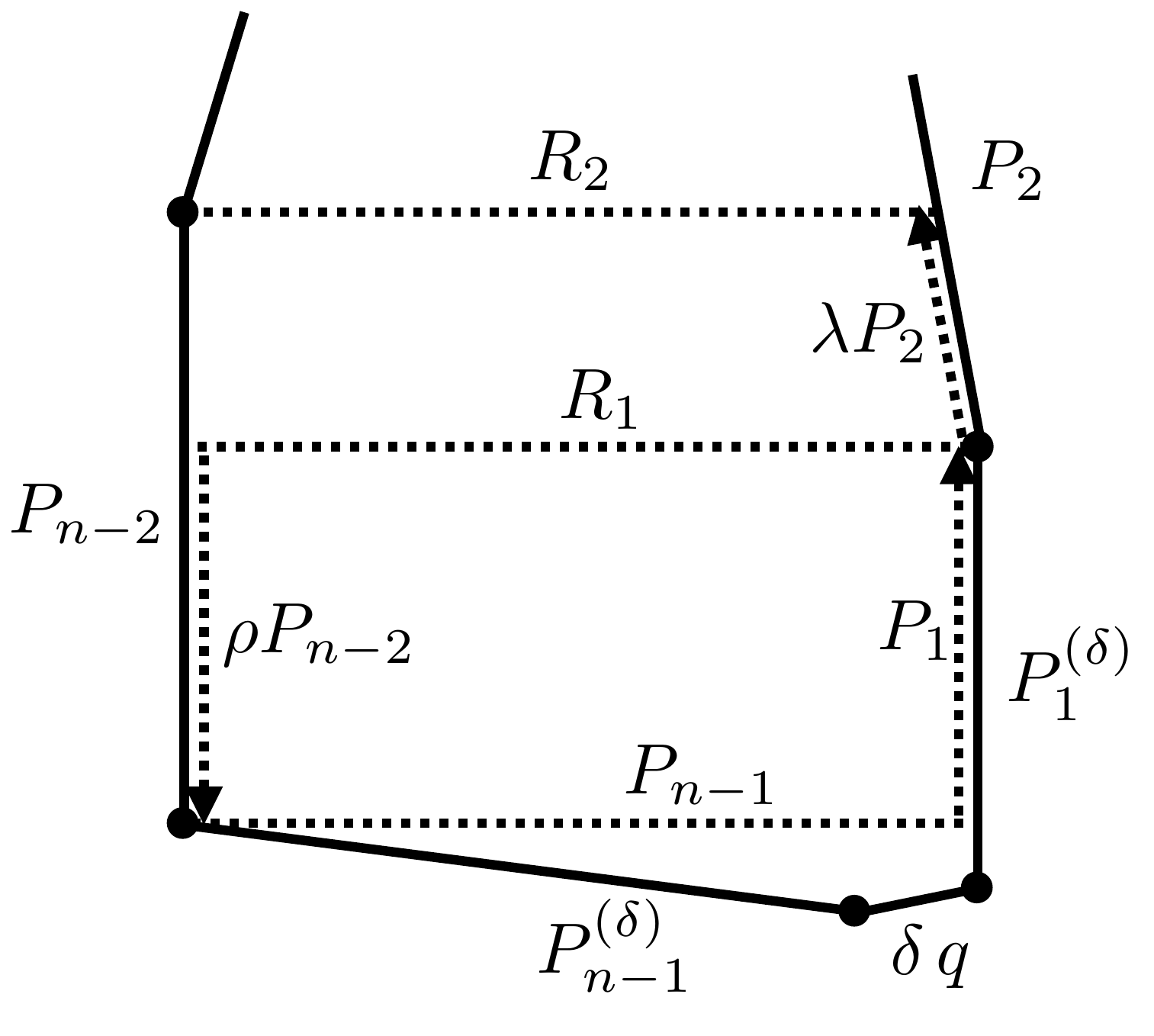}
\caption{We take the soft limit as shown, with $\delta \, q \to 0$ and $p_{n-1}^{(\delta)} \to p_{n-1},~p_{1}^{(\delta)} \to p_{1}$. $\delta$ is the small parameter in the soft limit.}
\label{figa2}
\end{figure}

We now implement the construction we have just discussed in terms of specific momenta. We choose the null momenta as shown in figure \ref{figa2}. The soft limit will be implemented by taking $\delta \to 0$. 
There are many polygons in play, which we now describe in detail. 
\begin{itemize}
\item The original polygon ${\cal W}_n$. This is composed of the null momenta 
$$\cdots, p_{n-2}, p_{n-1}^{(\delta)},\delta\,q,p_1^{(\delta)}, p_2,\cdots$$
\item The polygon ${\cal W}_{n-1}$ after we take the soft limit. In the soft limit $p_{n-1}^{(\delta)} \to p_{n-1}$ and $p_{1}^{(\delta)} \to p_{1}$, so the polygon ${\cal W}_{n-1}$ is composed of the null momenta
$$\cdots, p_{n-2}, p_{n-1},p_1, p_2,\cdots$$
\item The reference square ${\cal W}^{(1)}_4$ formed by $p_{n-1},p_1,R_1,\rho \, p_{n-2}$. The WL OPE is defined in terms of symmetries of this reference square. Here $\rho$ is a proportionality factor. It can be computed from momentum conservation together with the fact that $R_1$ is null, so that
$$(\rho \, p_{n-2}+p_{n-1}+p_1)^2=0$$
\item The reference pentagon ${\cal W}^{(1)}_5$ formed by $R_1,\rho\, p_{n-2},p_{n-1}^{(\delta)},\delta \, q,p_1^{(\delta)}$, which is used in the definition of the ratio function $\hat {\cal R}$. 
\item The reference hexagon ${\cal W}^{(1)}_6$ formed by $p_{n-2}, p_{n-1}^{(\delta)}, \delta \, q,p_1^{(\delta)}, \lambda p_2$, and $R_2$, whose cross-ratios control the WL OPE. 
\end{itemize}
The superscripts on ${\cal W}^{(1)}_{4,5,6}$ remind us that we can define several reference squares, pentagons, and hexagons along the tessellation of ${\cal W}_n$. Only the ones described here will be relevant for the leading-log corrections. We assume that $q$ is null, consider generic kinematics, so that denominators below do not vanish, and choose the collinear prescription $p_1^{(\delta)} \parallel p_1$. We take the soft limit such that 
\begin{equation}
p_{n-1}^{(\delta)}+\delta \, q+p_{1}^{(\delta)} = p_{n-1}+p_1,
\end{equation}
so that momentum is conserved throughout the process. Given that $p_{n-1}^{(\delta)}$ is null, we get, at leading order in $\delta$,
\begin{equation}
p_{1}^{(\delta)}= p_1 -\delta \frac{p_{n-1} \cdot q}{p_{n-1} \cdot p_1}  p_1.
\end{equation}
The reference hexagon is built from the momenta shown in the figure, in particular $\lambda p_2$. For the pentagon inside the hexagon, we have (all arrows point counter-clockwise as we move along the pentagon)
\begin{equation}
R_2+p_{n-2} + p_{n-1}+p_1+\lambda p_2=0.
\end{equation}
We can solve for $\lambda$ by requiring $R_2$ to be null. 
\begin{equation}
(p_{n-2} + p_{n-1}+p_1+\lambda p_2)^2=0
\end{equation}
which gives a linear expression for $\lambda$ written purely in terms of the momenta after taking the soft limit. 

We are now in a position to determine the dictionary. Following (\ref{hexagonratios}), we can write the cross-ratios of the reference hexagon in terms of the corresponding momenta. Using (\ref{hexagonparam}) then provides the map to $\tau,\sigma,\phi$. 
We obtain the following correspondence near the soft limit:
\begin{eqnarray}
2e^{-\sigma - \tau} { \cos \phi} &=& - \frac{p_1 \cdot p_2 + p_2 \cdot p_{n-1}}{p_1 \cdot p_2 ~ p_{n-1} \cdot p_1} \delta \, q \cdot p_1 - \frac{p_1 \cdot p_{n-2} + p_{n-2} \cdot p_{n-1}}{p_1 \cdot p_{n-1} ~ p_{n-2} \cdot p_{n-1}} \delta \, q \cdot p_{n-1} \nonumber \\
&& +\frac{\delta \, q \cdot p_2}{p_1 \cdot p_2} + \frac{\delta \, q \cdot p_{n-2}}{p_{n-2} \cdot p_{n-1}}+{\cal O}(\delta^2) 
\label{eq:Usoftdef}
\end{eqnarray}
We denote the displayed term linear in $\delta$ by $U(p_{n-2},p_{n-1},\delta \, q,p_1,p_2)$. It follows that $\delta \sim e^{-\sigma -\tau}$. We furthermore find
\begin{equation}
\sigma =-\frac{1}{2} \log \delta \, q \cdot p_1 + \cdots ,~~~\tau = -\frac{1}{2}\log \delta \, q \cdot p_{n-1}+ \cdots
\label{eq:sigma-tau-soft-map}
\end{equation}
Corrections to these relations can be disregarded in the limit of interest. We can now put all the pieces together and give the following prediction for the universal leading-log correction in the soft limit:
\begin{equation}
\log {\hat {\cal R}}_{n} - \log {\hat {\cal R}}_{n-1} = U(p_{n-2},p_{n-1},\delta \, q,p_1,p_2) P_{\rm LL}(-\frac{1}{2}\log \delta \, q \cdot p_1,-\frac{1}{2}\log \delta \, q \cdot p_{n-1}) + \cdots
\label{eq:Rhat-soft-LL}
\end{equation}
At one loop, this can be compared to the results of \cite{Brandhuber:2015vhm}. For this, we need to recall the definition of the ratio function
\begin{equation}
{\hat {\cal R}}_{n}(q) = \frac{{\cal W}_n(\delta \, q)}{{\cal W}_5^{(1)}(\delta \, q)} \times \text{(independent of $q$)} 
\end{equation}
and insert the one-loop logarithmic divergences found in \cite{Brandhuber:2015vhm} for ${\cal W}_n(q)$ and ${\cal W}_5^{(1)}(q)$. We find precise agreement. Starting at two loops, $\log {\cal W}_{n}$ agrees with the remainder function up to corrections proportional to the one-loop answer (times the cusp anomalous dimension). These are suppressed in the limit considered here. Hence
\begin{equation}
R^{(\ell)}_n  = R^{(\ell)}_{n-1} + U(p_{n-2},p_{n-1},\delta \, q,p_1,p_2) P_{\rm LL}^{(\ell)}(-\frac{1}{2}\log \delta \, q \cdot p_1,-\frac{1}{2}\log \delta \, q \cdot p_{n-1})
\label{eq:R-soft-LL}
\end{equation}

for $\ell=2,3,\cdots$.

\bibliographystyle{JHEP}
\bibliography{N4refs}

\end{document}